\documentclass[journal,twoside]{IEEEtran}
\pdfoutput=1
\usepackage{cite}
\usepackage[cmex10]{amsmath}
\usepackage{graphicx}
\graphicspath{{./}{figs/}}
\DeclareGraphicsExtensions{.pdf,.png}
\usepackage{array}
\usepackage{mathptmx}
\usepackage{mathtools}
\usepackage{amssymb}
\usepackage{nicefrac}
\usepackage{enumitem}
\usepackage{booktabs}
\usepackage{subfigure}
\usepackage{url}
\usepackage{trfsigns}
\usepackage[OT2,T1]{fontenc}
\usepackage{color}

\newcommand{\abs}[1]{{\left\lvert#1\right\rvert}}
\newcommand{\rect}{\operatorname{rect}}
\newcommand{\sinc}{\operatorname{sinc}}
\newcommand{\sgn}{\operatorname{sgn}}

\definecolor{blue}{rgb}{0,0,0}
\definecolor{red}{rgb}{0,0,0}
\definecolor{magenta}{rgb}{0,0,0}

\usepackage{dblfloatfix}

\begin{document}

\title{Decision-Based Transcription of Jazz Guitar Solos Using a Harmonic Bident Analysis Filter Bank and Spectral Distribution Weighting}

\author{Stanislaw~Gorlow,~\IEEEmembership{Member,~IEEE,} Mathieu~Ramona, and~Fran\c{c}ois~Pachet%
\thanks{This work was financially supported by the European Commission under grant EU-FP7-318770.}%
\thanks{S. Gorlow was with Sony Computer Science Laboratory (CSL) in Paris, 6 rue Amyot, 75005 Paris, France. He is now with Dolby Sweden.}
\thanks{M. Ramona and F. Pachet are with Sony CSL Paris.}%
\thanks{This paper has supplementary downloadable material. The material includes jazz guitar solo excerpts (licks) and their respective transcriptions. Contact stanislaw.gorlow@gmail.com for further questions about this work.}}



\maketitle

\begin{abstract}
Jazz guitar solos are improvised melody lines played on one instrument on top of a chordal accompaniment (comping). As the improvisation happens spontaneously, a reference score is non-existent, only a lead sheet. There are situations, however, when one would like to have the original melody lines in the form of notated music, see the Real Book. The motivation is either for the purpose of practice and imitation or for musical analysis. In this work, an automatic transcriber for jazz guitar solos is developed. It resorts to a very intuitive representation of tonal music signals: the pitchgram. No instrument-specific modeling is involved, so the transcriber should be applicable to other pitched instruments as well. Neither is there the need to learn any note profiles prior to or during the transcription. Essentially, the proposed transcriber is a decision tree, thus a classifier, with a depth of 3. It has a (very) low computational complexity and can be run on-line. The decision rules can be refined or extended with no or little musical education. The transcriber's performance is evaluated on a set of ten \textcolor{red}{jazz} solo excerpts and compared with a state-of-the-art transcription \textcolor{blue}{system} for the guitar \textcolor{red}{plus PYIN}. \textcolor{blue}{We achieve} an improvement of 34 \% \textcolor{red}{w.r.t.\ the reference system and 19 \% w.r.t.\ PYIN} in terms of the F-measure. Another measure of accuracy, the error score, attests that the number of erroneous pitch detections is reduced by more than 50 \% \textcolor{red}{w.r.t.\ the reference system and by 45 \% w.r.t.\ PYIN}. 
\nocite{Leonard2005}
\end{abstract}



\IEEEpeerreviewmaketitle

\section{Introduction}

\IEEEPARstart{A}{utomatic} music transcription, or audio-to-score conversion, is for many the ``Holy Grail'' of music informatics thanks to the apparent complexity of the task. The problem is even harder in the presence of multiple melodic voices. Over the years, a plethora of different approaches has been proposed; too many to name them all. Some worthwhile ideas can be found in \cite{Hermes1988,Goto2004,Klapuri2008,Yeh2010,Dressler2011,Wu2011} and in the references given therein. For monophonic signals, it can be said that YIN \cite{Cheveigne2002} has established itself as \textcolor{blue}{a standard reference}. For the more complex polyphonic signals, where multiple notes may be played by different instruments at the same time, any form of a nonnegative matrix factorization (NMF) won the popular vote; the NMF is usually applied to the spectrogram or its square root \cite{Smaragdis2003}. A more recent comparison of some of the state-of-the-art techniques was made in \cite{Berg2014_NIPS} on the basis of piano signals. At first glance, the winning algorithm looks a lot like \cite{Fuentes2013_ASL}, which then again looks like a probabilistic reformulation of the model in \cite{Vincent2010} including a noise term. 

As of today, one of the more popular tasks in music information retrieval constitutes the detection of the dominant melodic voice in a homophonic musical texture \cite{Salamon2014}. Many times, however, a music signal in which two or more notes are played simultaneously is referred to as \emph{polyphonic}---without further looking into the semantics. Melody detection has its roots in pitch detection of human voice, which actually means the detection of the fundamental frequency of a harmonic oscillator. As mentioned earlier, the most prominent pitch detection algorithm today is YIN. It was originally designed for speech, but it is generic enough to be applied to (monophonic) music as well. In the context of speech, one may also mention \cite{Zahorian2008}. In their work, the authors manage to outperform YIN by a narrow margin with a much more complex and finely tuned overall system. \textcolor{red}{A more recent variant of a YIN-based pitch-tracking system is PYIN \cite{Mauch2014}.} A longer list of algorithms to perform (monophonic) melody extraction from a polyphonic music signal can be found in \cite{Salamon2014}. 

Generally, it can be said that \textcolor{blue}{neither YIN nor PYIN} perform well if not at all on polyphonic signals. NMF-like approaches are non-musical by design, and as such they often generate non-musical artifacts that require manual intervention or any sort of musical post-processing. Probabilistic approaches become computationally intractable if musical constraints in the form of additional priors are incorporated into them; one will have a hard time trying to find the corresponding formulation that must be both: mathematically \emph{and} musically sound. In many cases, the two objectives are opposed and cannot be achieved simultaneously. For these (and many other reasons), the most recent work that directly relates to this paper's topic pursues \textcolor{blue}{a} bag-of-features approach to reach a result \cite{Kehling2014_DAFx}. 

The greater part of pitch and melody detection algorithms relies on a score function, which is also termed a salience function to underline the perceptual component of pitch and melody, see \cite{Terhardt1982}. One major drawback of the most commonly used score functions is the appearance of phantom pitch at integer multiples of the fundamental, which as a consequence gives rise to octave errors. Another issue with present-day algorithms, which are capable of polyphony, is that they are trained mostly on piano recordings that miss the subtle nuances of a real performance. If you listen to a jazz guitar recording, apart from picked or plucked notes and chords you will notice arpeggios, hammer-ons and hammer-offs, pitch modulations, fretting noise, etc. Put simply, all the elements that render not only a guitar performance expressive and vivid.

\textcolor{red}{The present work was mainly motivated by the fact that an accessible system for guitar performance analysis, which makes use of YIN in combination with a transient detector at its core, did not provide the desired precision and accuracy, in particular when applied to solo improvisations performed on an electric jazz guitar. Another drawback of the system was that it could not be run on-line, i.e.\ simultaneously with the actual performance. Hence, we decided to review the problem of pitch detection and pitch representation.}

In \textcolor{blue}{acoustic} music, the vast majority of playing techniques can be associated with distinct acoustical phenomena that relate to the physical properties of the instrument \cite{Benade1990,Meyer2009}. The number of these techniques is finite and can be reduced further if the context such as the artist or the genre is known. Hence, on condition that we possess a suitable signal representation to measure the concomitant phenomena, we can define a set of rules to identify the techniques employed by the performer and decipher the musical message. This is one of the central ideas behind the present paper. 

The spectrogram has emerged as the standard representation for speech (also called a \emph{voicegram}), music, and other signals that exhibit a structure along the frequency axis, which may also change over time \cite{Boashash2003}. However, for tonal music in particular, the key question is often not about the physical composition of the signal in terms of frequency components, but rather about its musical composition in terms of pitch or melody. And so, this work is not so much about pitch or melody detection as it is about finding a better representation for the harmonic content of a music signal, which as a natural consequence would facilitate the analysis and the transcription task. 

\textcolor{red}{The major contribution of the present work consists in the proposed filter bank.} The developed time-pitch transform may be viewed as a multi-pitch score function that varies over time. This type of representation we call a \emph{pitchgram}. A genuine pitchgram can assist a trained musician during transcription tasks or it can help an amateur find the melody, the solo, or the progression of chords in the accompaniment. A more reliable score function should also improve the accuracy of the salience-based detection algorithms mentioned in \cite{Salamon2014}. \textcolor{red}{In addition, we make public the used data set with the manual transcriptions for everybody interested in this area of research.}

To substantiate our \textcolor{blue}{statements in the preceding paragraphs}, we manually construct a decision tree out of empirical decision rules that translate into musical acoustics. The \textcolor{blue}{tree serves} as a pitch classifier, i.e.\ an automatic transcriber, while resorting to any of the two variants of the newly developed time-pitch representation for tonal signals as input. 
\textcolor{red}{We demonstrate that our new time-pitch representation is more reliable than, e.g., the (note) activation diagram of an NMF. On a final note, we would like to mention some similarities and differences w.r.t.\ the so-called ``specmurt'' analysis \cite{Saito2008}. Both methods do not require prior knowledge of the number of latent components, i.e.\ distinct pitches. Both are sensitive to signal amplitude: in the presence of strong components, weak components vanish due to normalization and thresholding.  The advantages of the filter bank approach with distribution weighting are:
\begin{itemize}
\item The pitch axis can be neither linear nor logarithmic;
\item Each harmonic series is weighted differently, i.e.\ there is no need to compute a common harmonic weighting; 
\item There are no instability issues related to inverse filtering or deconvolution and there are no iterations; 
\item The computation of the pitchgram is less costly, since it requires only a single (forward) Fourier transform in the case of the faster variant. 
\end{itemize}}

\textcolor{red}{The structure of the paper is as follows. In Section~\ref{sec:pitch}, the definition of pitch on a twelve-tone equal temperament scale is briefly reviewed in order to formulate the properties of the corresponding frequency spectrum. Inspired by the physics, a multi-tone pitch analysis filter bank is designed in Section~\ref{sec:filter_bank}. 
The output from the filter bank is a time-pitch representation of the signal: the pitchgram. The latter is used as input for a rule-based pitch classifier, which is detailed in Section~\ref{sec:transcription}. In the following Section~\ref{sec:evaluation}, the pitch classifier is then evaluated and compared. Section~\ref{sec:conclusion} concludes the paper.} 

\section{Pitch on a Twelve-Tone Equal Temperament Scale}
\label{sec:pitch}

\subsection{Pitch and Harmonic Spectrum}

Pitched musical instruments produce sounds, or notes, of definite pitch that can be discerned by a listener. Sounds with definite pitch usually possess a harmonic spectrum with many or only a few components. This is because a musical tone that is generated on a pitched instrument has several modes of vibration that occur simultaneously. \textcolor{red}{In steady state,} \textcolor{blue}{a} pitched instrument can hence be approximated as a harmonic oscillator, such as a string, which vibrates at the frequencies of a harmonic series.\textcolor{red}{\footnote{\textcolor{red}{Some musical instruments exhibit inharmonicity. However, inharmonicity is negligible for the nylon strings of guitars. It only becomes appreciable for thick steel strings, as in the piano \cite{Fletcher1999}.}}} The listener hears all frequencies at once. The frequency of the slowest vibration is the fundamental frequency, which is usually perceived as the musical pitch. The other frequencies are overtones. A harmonic series consists of frequencies that are integer multiples of the fundamental frequency, which is equally the first harmonic. The overtones start at the second harmonic. Harmonics are also called partials, referring to the different parts of the harmonic spectrum of a musical tone. Harmonics have the property that they are periodic at the fundamental frequency, and so is their sum. In what follows, the pitch is quantified as the fundamental frequency of a harmonic series, even though it is a subjective psychoacoustical attribute rather than a purely objective physical property \cite{Plack2006}.

\subsection{Equal-Tempered Chromatic Scale}

The chromatic scale is a musical scale with twelve tones or pitches \cite{Barbour2004}. On an \emph{equal-tempered} instrument, all the semitones, i.e.\ the intervals between two adjacent tones, have the same size. As a consequence, the notes of an equal-tempered chromatic scale are equally spaced. More precisely, the twelve-tone equal temperament divides the octave, which is the interval between two consecutive harmonics, into twelve segments, all of which are equal on a logarithmic scale. Equal-temperament tuning systems are widely used in contemporary music. The MIDI tuning standard, e.g., by default expresses pitch on that scale.\footnote{See \url{http://www.midi.org/}}

\section{Pitch-Synchronous Chromatic Bident Analysis Filter Bank}
\label{sec:filter_bank}

\begin{figure*}[!t]
\centering
\includegraphics[width=0.8\textwidth]{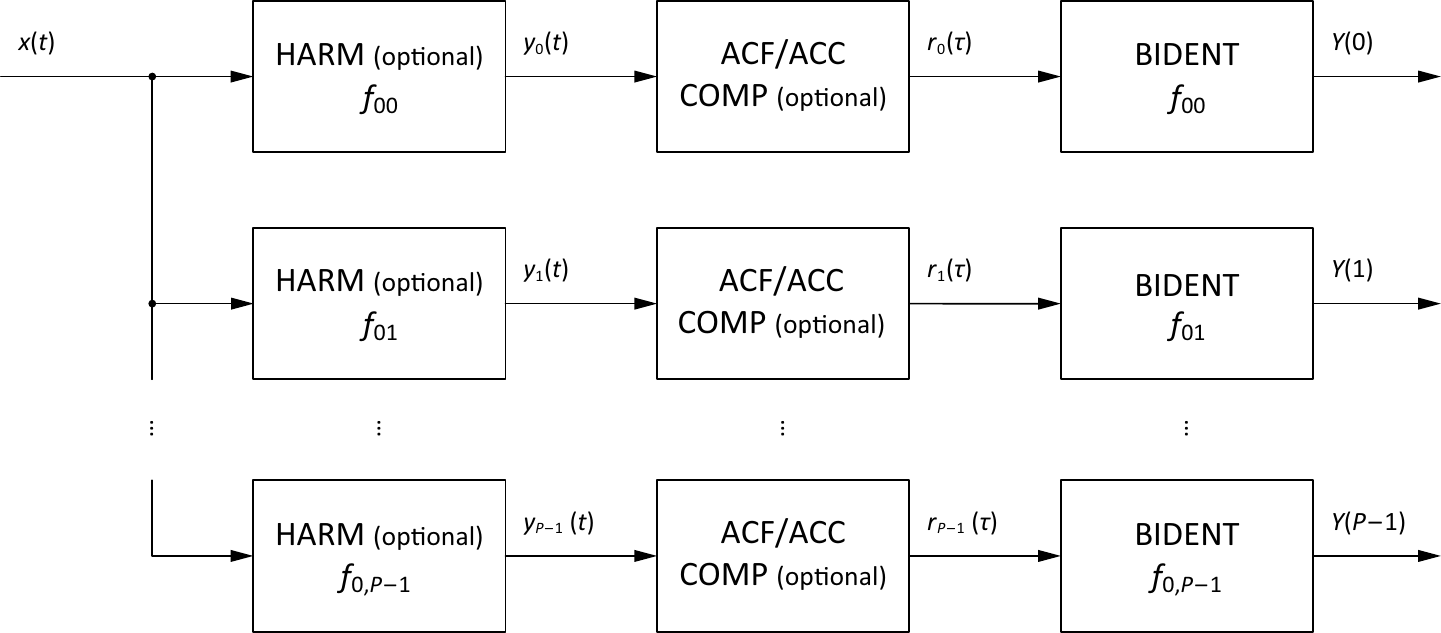}
\caption{$P$-tone pitch-synchronous chromatic bident analysis filter bank.}
\label{fig:filterbank}
\end{figure*}

In this section, a \textcolor{blue}{novel} pitch-analysis filter bank is \textcolor{blue}{designed}. It has for purpose the decomposition of a \textcolor{blue}{(tonal)} signal into a set of pitch coefficients, each one being associated with a distinct tone or note. \textcolor{red}{In principle, t}he signal's texture can be monophonic, polyphonic or homophonic\textcolor{red}{, albeit the latter case is treated only marginally}. 
\textcolor{red}{The filter bank is shown in Fig.~\ref{fig:filterbank}.} \textcolor{blue}{Its} constituent blocks are discussed in the \textcolor{blue}{following} sections. 
The particularity of the proposed filter bank lies in its ability to alleviate octave errors \textcolor{red}{on the hypothesis that a tone spectrum is harmonic and has a decaying tendency}, which can facilitate further analysis. 

\subsection{Pitch-Synchronous Comb Filter and Harmonicity Coefficient}

\subsubsection{Pitch-Synchronous Comb Filter}

\begin{figure}[!t]
\centering
\ifCLASSOPTIONonecolumn
\includegraphics[width=.9\columnwidth]{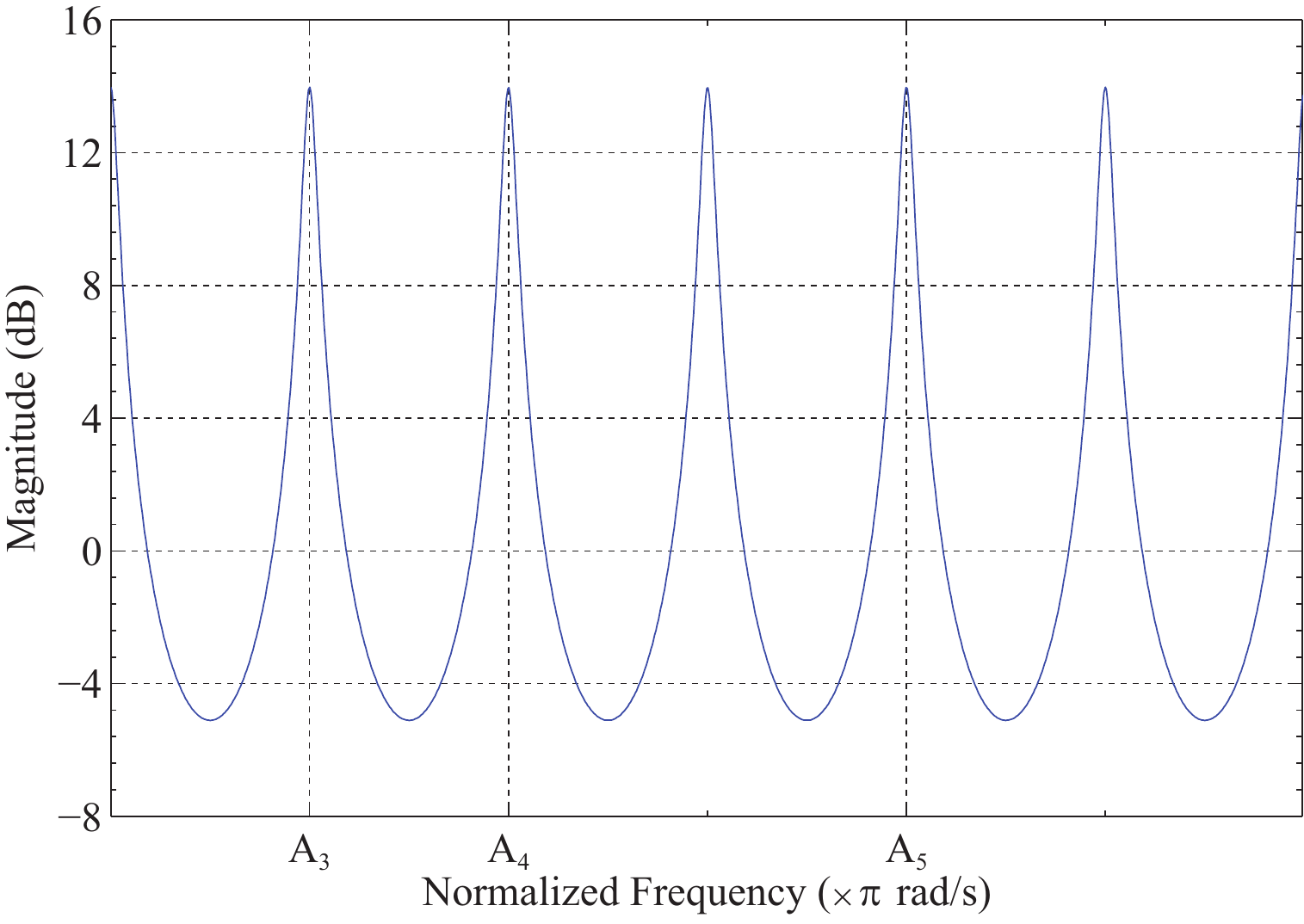}
\else
\includegraphics[width=\columnwidth]{combfa3}
\fi
\caption{Feed-backward comb filter response for A\textsubscript{3} ($a = 0.8$).}
\label{fig:comb}
\end{figure}

Consider the relation between a tonal input signal $x(t)$ and the corresponding output signal $\tilde{x}(t)$ to be defined as \cite{Smith2010}
\begin{equation}
\tilde{x}(t) = x(t) + a \, x{\left(t - T_0\right)} \text{,}
\label{eq:forward}
\end{equation}
where $T_0$ is the pitch period and $a \in {\left[-1, 1\right]}$ is a scaling factor. The difference equation \eqref{eq:forward} describes the \emph{feed-forward} comb filter. Its frequency response consists of a series of regularly spaces \textcolor{red}{notches}, hence the name. It is based on the principle of constructive and destructive interference between two waves that are correlated or coherent with each other. The alternative \emph{feed-backward} comb filter is described by 
\textcolor{red}{\begin{equation}
\tilde{x}(t) = x(t) + a \, \tilde{x}{\left(t - T_0\right)} \text{.}
\label{eq:backward}
\end{equation}}%
Note that the feed-backward comb filter variant is stable only if $\abs{a} < 1$. The magnitude response of the filter for a delay equal to the pitch period of A\textsubscript{3} is shown in Fig.~\ref{fig:comb}. It can be seen that the spikes appear at integer multiples of the fundamental including the zero frequency (direct current). Each channel of the analysis filter bank has an optional comb filter at the entrance that resonates at the nominal frequency and the corresponding integer multiples. Accordingly, the nominal frequency in Fig.~\ref{fig:comb} is the frequency of A\textsubscript{3}, which is 220 Hz. The purpose of the pre-filtering is to accentuate the components of a complex tonal mixture at the frequencies of a particular harmonic series and to attenuate all other components. \textcolor{red}{In the remaining part of the manuscript, we resort to the feed-backward variant of the comb filter.} 

\subsubsection{Harmonicity Coefficient}
\label{sec:harmonicity}

To assess the presence of a particular harmonic series in the input signal $x(t)$, we introduce a \emph{harmonicity coefficient}. We define it as the root mean square (RMS) of (desirably) one period of the comb-filtered signal $\tilde{x}(t)$, 
\begin{equation}
\eta\textcolor{red}{(T_0)} \triangleq \sqrt{\frac{1}{T_0} \int_0^{T_0}{{\left[\tilde{x}(t)\right]}^2 \, \mathrm{d} t}} \text{.} 
\end{equation}
The output of the first block is the original input signal $x(t)$ weighted by its harmonicity coefficient w.r.t.\ the nominal frequency of the respective filter bank channel, not the filtered signal: 
\begin{equation}
y(t\textcolor{red}{, T_0}) = \eta\textcolor{red}{(T_0)} \, x(t) \text{.} 
\label{eq:weighting}
\end{equation}

\subsection{Autocorrelation Function and Compression}
\label{sec:acf}

\subsubsection{Autocorrelation Function}

The second block initially computes the autocorrelation function (ACF) of the pre-weighted signal $y(t)$, which consists of multiple decaying complex tones with distinct harmonic spectra. The ACF is typically defined as the limit of the time average, 
\begin{equation}
{r_{yy}(\tau) = {\lim_{T \, \rightarrow \, \infty}} \, {\frac{1}{T}} {{\int_0^T} {y(t + \tau) \, y(t)} \, \mathrm{d} t} \text{,}}
\label{eq:acf}
\end{equation}
where $\tau$ denotes the time lag and $T$ is the observation window, respectively. The mean signal power is given by the ACF at zero lag, i.e.
\begin{equation}
r_{yy}(0) = {\lim_{T \, \rightarrow \, \infty}} \, {\frac{1}{T}} {{\int_0^T} {{\left[y(t)\right]}^2} \, \mathrm{d} t} \text{.}
\label{eq:pow}
\end{equation}
When normalized by the mean signal power, the autocorrelation function is often referred to as the autocorrelation coefficient (ACC) 
\begin{equation}
\rho_{yy}(\tau) = \frac{r_{yy}(\tau)}{r_{yy}(0)} \text{.}
\label{eq:acc}
\end{equation}
Should the ACC and not the ACF be computed in the second block, the harmonic weighting can be skipped. This is why \eqref{eq:weighting} is optional. Two important properties of the ACF (and the ACC) are that it is (a) \textcolor{blue}{real-valued} and (b) evenly symmetric. Hence, its Fourier transform is \textcolor{blue}{real-valued} and evenly symmetric as well. These properties are exploited in the next section, where the chromatic bident filter is derived. 

\subsubsection{Compression}
\label{sec:compression}

In quite a few papers on polyphonic music transcription, see e.g.\ \cite{Gorlow2015_LVA}, it was observed that often better results are obtained if the magnitude spectrum is used instead of the power spectrum along the frequency dimension. And since the power spectrum is the Fourier transform of the ACF, we achieve a similar effect by taking the square root of $\abs{r_{yy}(\tau)}$ while keeping the sign, i.e.\ 
\begin{equation}
\tilde{r}_{yy}(\tau) \triangleq {\sgn}{\left[r_{yy}(\tau)\right]} \, \abs{r_{yy}(\tau)}^{\nicefrac{1}{2}} \text{,} 
\end{equation}
where $\sgn$ denotes the signum or sign function. We call this \emph{compression}. 

\subsection{Cosine-Modulated Chromatic Bident Filter}
\label{sec:bident}

\subsubsection{Cosine-Modulated Chromatic Sinc Filter}

Imagine that we wanted to design a filter that has its peak or maximum where is the fundamental frequency of one particular tone and zeros at the fundamental frequencies of all other tones in a given range. An ideal filter that fits the description is a shifted Dirac delta function. However, since its impulse response, which is equivalent to the size of the observation window, is infinite, the Dirac delta filter is impracticable. If we truncate the window, the Dirac delta becomes the sinus cardinalis or $\sinc$ for short. The width of the main lobe of the $\sinc$ function is inversely proportional to the window size, while the zero crossings of the normalized $\sinc$ occur at non-zero integers; the zero crossings are \emph{equidistant}. Therefore, if we assume that each tone in a tonal mixture carries most of its energy in the harmonic partials, we can dimension a rectangular window in such a manner that the first zero crossing is \emph{one twelfth} of an octave away from the fundamental frequency of the tone of interest in the lower octave, i.e.\ between $\nicefrac{f_0}{2}$ and $f_0$; one twelfth because we are considering a chromatic scale. This is to prevent lower-pitched and also higher-pitched tones from leaking significantly into the filtered signal, which ideally would oscillate at its fundamental frequency only. Note that the number of zero crossing in the upper octave, i.e.\ between $f_0$ and $2\,f_0$, is \emph{twice} as large. Attention should also be paid to the fact that the leakage from neighboring tones is not perfectly suppressed. This is because the zero crossings of the $\sinc$ filter are equidistant on a linear scale, whereas the pitches are equidistant on a logarithmic scale. 
The effect is negligible, however, as the side lobes of the $\sinc$ decay rather quickly. This is further supported by the fact that most of the energy is concentrated under the main lobe with the area under the side lobes adding up to a small (negative) value. 


The window discussed above is formally given by the rectangle function, which is defined as
\begin{equation}
{\rect}{\left(\frac{t}{24 \, T_0}\right)} = 
	\begin{dcases}
		1 & \text{if}\ \abs{t} \leqslant 12 \, T_0 \text{,} \\
		0 & \text{otherwise,}
	\end{dcases}
	\label{eq:rect}
\end{equation}
where $T_0$ is the pitch period of the tone of interest. Accordingly, its (unitary) Fourier transform is given by \cite{Oppenheim2009}: 
\begin{equation}
{\rect}{\left(\frac{t}{24 \, T_0}\right)} \ \fourier 24 \, T_0 \, {\sinc}{\left(24 \, \frac{f}{f_0}\right)} 
\end{equation}
with
\begin{equation}
{\sinc}{\left(24 \, \frac{f}{f_0}\right)} = \frac{{\sin}{\left(24 \, \pi \, \nicefrac{f}{f_0}\right)}}{24 \, \pi \, \nicefrac{f}{f_0}} \text{\textcolor{red}{,}}
\label{eq:sinc}
\end{equation}
\textcolor{red}{where $\fourier$ symbolizes the direction of the mapping: from a \emph{discrete} (time) domain $\circ$ to a \emph{continuous} (frequency) domain.} From \eqref{eq:sinc} we can immediately see that the $\sinc$ has twelve zero crossings in the lower octave and twenty-four of them in the upper octave, respectively. In the previous section it is underlined that the Fourier transform of the ACF is real. Now, if we window \eqref{eq:acf} by \eqref{eq:rect} and then apply the Fourier transform, we obtain a normalized cosine transform at $f_0$, which again is real, 
\begin{equation}
\tilde{Y}{\left(f_0\right)} \triangleq\ \frac{1}{24 T_0} \, {\int_{-\infty}^\infty} {{\rect}{\left(\frac{\tau}{24 T_0}\right)} \, r_{yy}{\left(\tau\right)}} \, {\cos}{\left(2 \pi \, f_0 \, \tau\right)} \, {\mathrm{d}} \tau \text{.}
\label{eq:cos_transform}
\end{equation}
The normalization term before the integral makes sure that the integral is \emph{independent} of the window size, and so the pitch period $T_0$. Note that \eqref{eq:cos_transform} is tantamount to a \emph{zero-lag cross-correlation} between the cosine-modulated rectangle window and the ACF of the comb-filtered input signal $x(t)$. But what is more: because the rectangle, the ACF, and the cosine are all even functions, the cross-correlation between the above terms is identical with their \emph{convolution}. Therefore, \eqref{eq:cos_transform} can be interpreted as a \emph{filtering operation} on the Fourier transform of the ACF, i.e. the harmonically weighted input power spectrum, by a pitch-shifted $\sinc$. In the frequency domain, this corresponds to a multiplication of the signal spectrum by the sinc function. Finally, due to symmetry, \eqref{eq:cos_transform} simplifies to
\begin{equation}
\tilde{Y}{\left(f_0\right)} = \frac{1}{12 T_0} \, {\int_0^{12 T_0}} {r_{yy}{\left(\tau\right)} \, {\cos}{\left(2 \pi \, f_0 \, \tau\right)} \, {\mathrm{d}} \tau} \text{.}
\end{equation}

\subsubsection{Cosine-Modulated Bident Window}

In the previous paragraph it was shown how a particular partial from a harmonic series can be filtered out, while suppressing simultaneously the neighboring tones. But is there a means to demarcate an overtone from the fundamental frequency of the same harmonic series? Or in other words, how can we \emph{score high} the \emph{first partial}, while \emph{scoring low} all the \emph{other partials} at the same time; and to make it more difficult, by the use of a simple filter?

\begin{figure}[t]
\subfiguretopcaptrue
\centering
\ifCLASSOPTIONonecolumn
\subfigure[Time domain]{\includegraphics[width=.9\columnwidth]{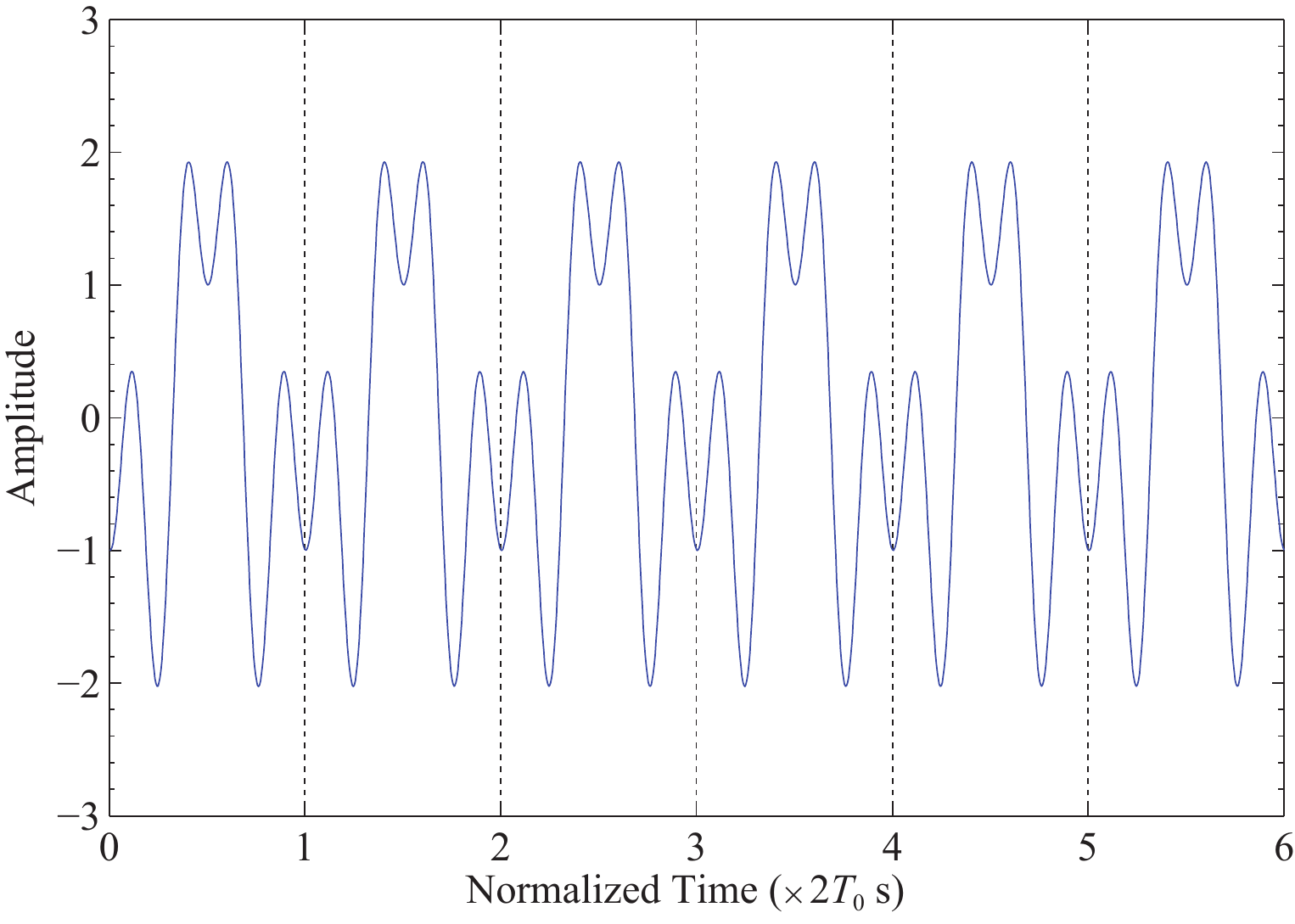}\label{fig:bident_time}}
\else
\subfigure[Time domain]{\includegraphics[width=\columnwidth]{concerta01}\label{fig:bident_time}}
\fi
\\
\ifCLASSOPTIONonecolumn
\subfigure[Frequency domain]{\includegraphics[width=.9\columnwidth]{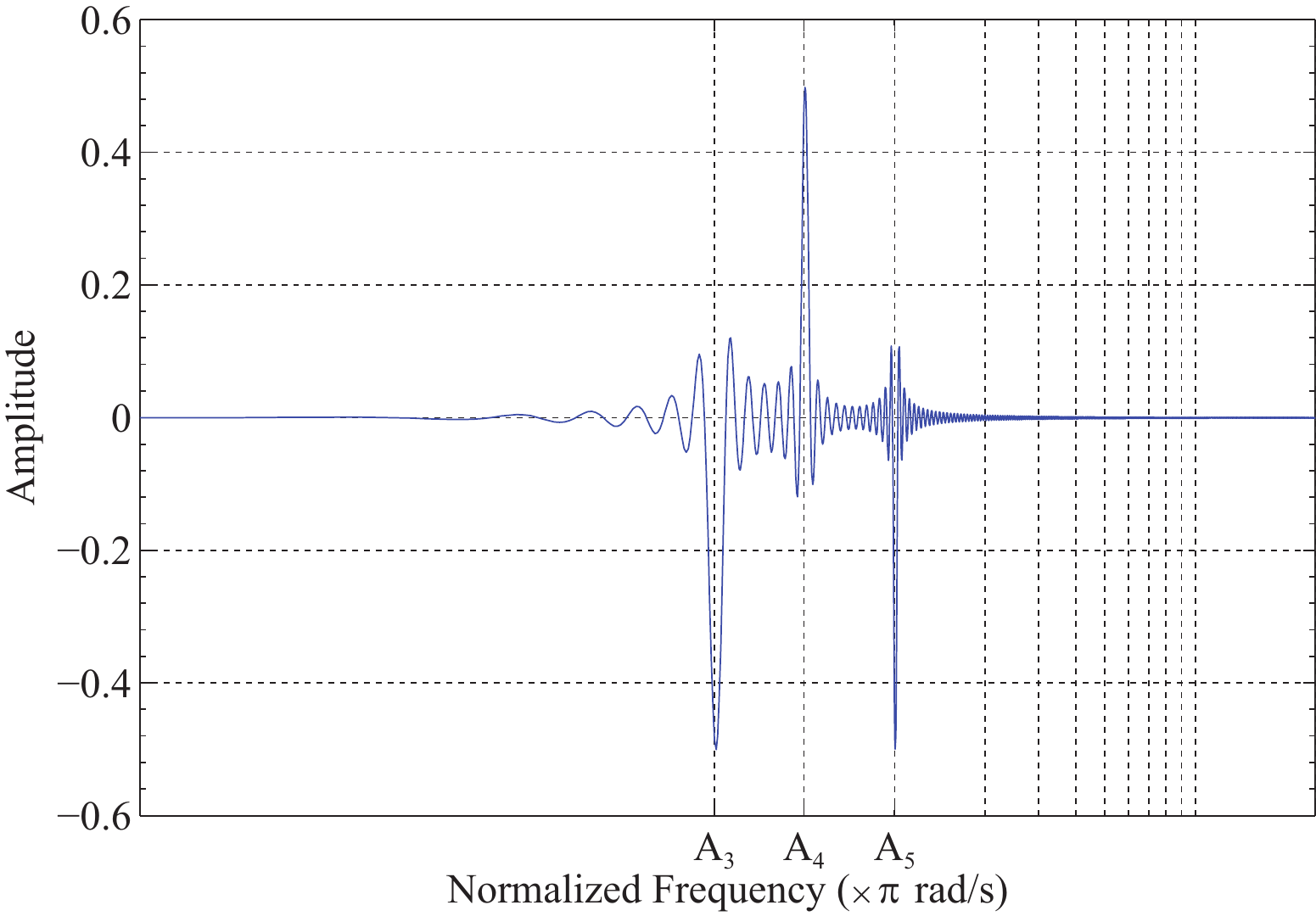}\label{fig:bident_freq}}
\else
\subfigure[Frequency domain]{\includegraphics[width=\columnwidth]{concerta02}\label{fig:bident_freq}}
\fi
\caption{(a) Cosine-modulated bident window and (b) the corresponding filter response for A\textsubscript{4}.}
\label{fig:bident}
\end{figure}

Consider the following prototype function:
\begin{equation}
{g(t) = \alpha \, {\sin}{\left(3 \pi \, f_0 \, t\right)} \, {\tan}{\left(\pi \, f_0 \, t\right)} - \beta \text{.}}
\label{eq:window}
\end{equation}
If we modulate the zero-frequency carrier in \eqref{eq:window} by half the fundamental frequency of a tone, we obtain
\begin{equation}
{h(t) \triangleq g(t) \, {\cos}{\left(\pi \, f_0 \, t\right)} \text{,}}
\label{eq:iir}
\end{equation}
which is the \textcolor{blue}{periodic} (infinite) impulse response of our filter. Its frequency response is given by the Fourier transform, which is 
\begin{equation}
H(f) = \tfrac{\alpha}{4}\, {\left[{\delta}{\left(f - f_0\right)} - {\delta}{\left(f - 2 f_0\right)}\right]} - {\tfrac{\beta}{2}} \, {\delta}{\left(f - \nicefrac{f_0}{2}\right)} \text{,}
\end{equation}
consisting of three weighted Dirac deltas with different signs at the frequencies $\nicefrac{f_0}{2}$, $f_0$, and $2\,f_0$. If we multiply \eqref{eq:iir} by \eqref{eq:rect} to limit the length of the filter, it becomes
\begin{align}
&\alpha \, 6 \, T_0 \, {\left[{\sinc}{\left(24 \, \frac{f - f_0}{f_0}\right)} - {\sinc}{\left(24 \, \frac{f - 2 f_0}{f_0}\right)}\right]}\\
&\quad {} - \beta \, 12 \, T_0 \, {\sinc}{\left(24 \, \frac{f - \nicefrac{f_0}{2}}{f_0}\right)} \Fourier\ {\rect}{\left(\frac{t}{24 \, T_0}\right)} \, h(t) \text{\textcolor{red}{,}} \nonumber
\end{align}
\textcolor{red}{where $\Fourier$ now symbolizes the inverse mapping.} 
In the case where $\alpha = 2$ and $\beta = 1$, the three spikes have the exact same magnitude, see Fig.~\ref{fig:bident}. Due to the shape of the truncated window $w(t)$ in the frequency domain,
\begin{equation}
w(t) \triangleq {\rect}{\left(\frac{t}{24 T_0}\right)} g(t) \text{,}
\end{equation}
it is given the name \emph{harmonic bident}. From Fig.~\ref{fig:bident_freq} we see that between two adjacent spikes there are eleven zero-crossings in the lower octave and twenty-three in the upper octave. Accordingly, with the new window, \eqref{eq:cos_transform} becomes
\begin{equation}
Y{\left(f_0\right)} \triangleq \frac{1}{24 T_0} \, {\int_{-\infty}^{\infty}} {w(\tau)} \, {r_{yy}{\left(\tau\right)} \, {\cos}{\left(\pi \, f_0 \, \tau\right)} \, {\mathrm{d}} \tau} \text{,}
\label{eq:xcorr_long}
\end{equation}
or more compactly written
\begin{equation}
{{Y}{\left(f_0\right)} = \frac{1}{12 T_0} \, {\int_0^{12 T_0}} {r_{yy}{\left(\tau\right)} \, {h(\tau)} \, {\mathrm{d}} \tau}}
\label{eq:xcorr_short}
\end{equation}
with $T_0 = \nicefrac{1}{f_0}$.


The rationale behind the bident filter is the following. A pitched instrument produces a harmonic spectrum that is assumed to be \textcolor{blue}{(exponentially)} \emph{decaying}. \textcolor{red}{Broadly speaking, this is motivated by the second law of thermodynamics.} So, if the holder of the bident, i.e.\ the middle spike, is shifted to the pitch of a complex tone, the output of the filter contains all the energy of the fundamental, which is reduced by the energy of the first overtone (right spike). \textcolor{blue}{If} the spectrum has a decaying trend, the output is positive. However, if the holder is placed on any of the overtones, the filter output \textcolor{blue}{should be} negative, because the energy of the overtone is now reduced not only the the \emph{next} overtone but also by the \emph{previous} overtone (left spike), which in the case of the first overtone is the fundamental. Due to the falling trend of the harmonics, one may anticipate that \textcolor{blue}{most of} the overtones will have a \emph{negative} output and that the pitch will be the frequency with \textcolor{blue}{the largest} \emph{positive} correlation or \emph{score}. It should be noted, however, that the spectrum \textcolor{blue}{(of overtones)} may not always be strictly monotonically decreasing. This \textcolor{red}{can} \textcolor{blue}{depend} on the timbre of the instrument and the playing technique, which can put the accent on \textcolor{blue}{the first} or \textcolor{blue}{the first few} overtones rather than the fundamental. In that case, the fundamental might have a lower (positive) score than the accentuated \textcolor{blue}{overtones}. Then, it is up to the transcriber to select the lowest partial of a harmonic series with a significant score. \textcolor{red}{Any constructive interference effects between multiple harmonic series (polyphony) are not explained by this model. In \cite{Lienard2007}, e.g., the authors promote a somewhat similar idea: a comb with alternate polarity, which should minimize octave errors.} 

\subsection{Pitch Analysis of Non-Stationary Signals}

The extension of \eqref{eq:xcorr_long} or \eqref{eq:xcorr_short} to \textcolor{blue}{a succession} of tones is very much straightforward. You just choose the resolution of the time grid and apply \eqref{eq:xcorr_short} at all time instants for all possible values of $f_0$. Formally, this is stated as
\begin{equation}
{Y}{\left(m, f_0\right)} = \frac{1}{24 T_0} \, {\int_{-\infty}^{\infty}} {w(\tau)} \, {r_{yy}{\left(m, \tau\right)}} \, {\cos}{\left(\pi \, f_0 \, \tau\right)} \, {\mathrm{d}} \tau \text{,}
\end{equation}
or equivalently
\begin{equation}
{{Y}{\left(m, f_0\right)} = \frac{1}{12 T_0} \, {\int_0^{12 T_0}} {r_{yy}{\left(m, \tau\right)} \, {h(\tau)} \, {\mathrm{d}} \tau} \text{,}}
\label{eq:xcorr_grid}
\end{equation}
where $m \equiv m \, T_g$ \textcolor{red}{(equivalent to sampling)}, $m \in \mathbb{N}_0$\textcolor{red}{, and} 
\textcolor{red}{\begin{equation}
r_{yy}(m, \tau) = {\lim_{T \, \rightarrow \, \infty}} \, {\frac{1}{T}} {{\int_0^T} {y(m \, T_g + t + \tau) \, y(m \, T_g + t)} \, \mathrm{d} t} \text{.}
\end{equation}}%
$T_g$ denotes the resolution of the time grid. The resulting time-pitch diagram is what we call the \emph{pitchgram}. $T_g$ can be chosen as, e.g., the duration of a thirty-second note. It can be noted that a corresponding chromagram \cite{Wakefield1999} is easily obtained from \eqref{eq:xcorr_grid} by 
\textcolor{red}{\begin{equation}
Z(m, \zeta) = \sum\nolimits_{\mu \in \mathbb{N}^+}{Y{\left(m, \mu \cdot f_{0, \zeta}\right)}} \text{,} 
\end{equation}}%
where $\zeta$ denotes a \emph{pitch class} or chroma and $\mu$ is the \emph{octave number} or tone height. It should be of greater benefit to use the rectangle window, i.e.\ the $\sinc$ filter, in that case. Also note that $f_0 \in \mathbb{R}$ by definition, i.e.\ it may take on any value and is not restricted to any tuning standard. 

\subsection{Discrete-Time Discrete-Pitch Implementation}

For the realization of \eqref{eq:xcorr_grid} as a computer program, the time variable $t$ or $\tau$ must be discretized. But first, for the sake of convenience, the frequency variable $f_0$ is substituted for the MIDI note number $p$, i.e.
\textcolor{red}{\begin{equation}
Y{\left(p\right)} = Y{\left(440 \cdot 2^{\nicefrac{p \, - \, 69}{12}}\right)}
\label{eq:pitchmap}
\end{equation}}%
for $p = 0, 1, \dots, 127$ and $f_0$ in Hz. The MIDI note with the number $p = 69$ belongs to the concert A ($f_0=440$). The middle C lies nine semitones below ($p = 60$). Finally, the values of $Y(p)$ are evaluated at \textcolor{red}{$t = n \equiv n \, T_s$}, $n \in \mathbb{Z}$, where $T_s$ denotes the sampling period. The pitch period of a note $T_0$ is substituted for $N_0 \equiv N_0 \, T_s$, respectively. \textcolor{red}{It should be noted that the pitch mapping in \eqref{eq:pitchmap} is arbitrary and not binding. In other words, the pitch resolution can be finer than a semitone with any tuning other than 440 Hz for A and the pitch range can be limited to whatever makes sense w.r.t.\ the instrument.} 

With the above definitions, the feed-backward discrete-time comb filter reads
\textcolor{red}{\begin{equation}
\tilde{x}(n) = x(n) + a \, \tilde{x}{\left(n - N_0\right)} \text{,}
\end{equation}}%
and the harmonicity coefficient computes as 
\begin{equation}
\eta = \sqrt{\frac{1}{N_0} \sum_{n \, = \, 0}^{N_0 - 1}{{\left[\tilde{x}(n)\right]}^2}} \text{.} 
\end{equation}
Respectively, the discrete autocorrelation is defined as
\begin{equation}
r_{yy}(\nu) = {\lim_{N \, \rightarrow \, \infty}} \, {\frac{1}{N}} {{\sum_{n\,=\,0}^{N - 1}} {y(n) \, y(n - \nu)}}
\end{equation}
for $\nu = 0, 1, \dots, 12 N_0 - 1$. The cosine-modulated bident becomes 
\begin{equation}
{h}{\left(n\right)} = {g}{\left(n\right)} \, {\cos}{\left(\frac{\pi \, n}{N_0}\right)}
\end{equation}
with
\begin{equation}
g(n) = \alpha \, {\sin}{\left(\frac{3 \pi \, n}{N_0}\right)} \, {\tan}{\left(\frac{\pi \, n}{N_0}\right)} - \beta
\end{equation}
for $n = 0, 1, \dots, 12\,N_0 - 1$. Taking everything into account, the time-dependent score for the $m$th tone is given by
\begin{equation}
{{Y}{\left(m, p\right)} = \frac{1}{12 N_0} \, {\sum_{n\,=\,0}^{12 N_0 - 1}} {r_{yy}{\left(m, n\right)} \, {h(n)}} \text{.}}
\label{eq:pitchgram}
\end{equation}
\textcolor{red}{The corresponding MATLAB code is part of the downloadable material.}

%

\subsection{Example}
\label{sec:example}

\begin{figure*}[!t]
\subfiguretopcaptrue
\centering
\subfigure[Spectrogram (logarithmic)]{
	\includegraphics[height=.35\textwidth]{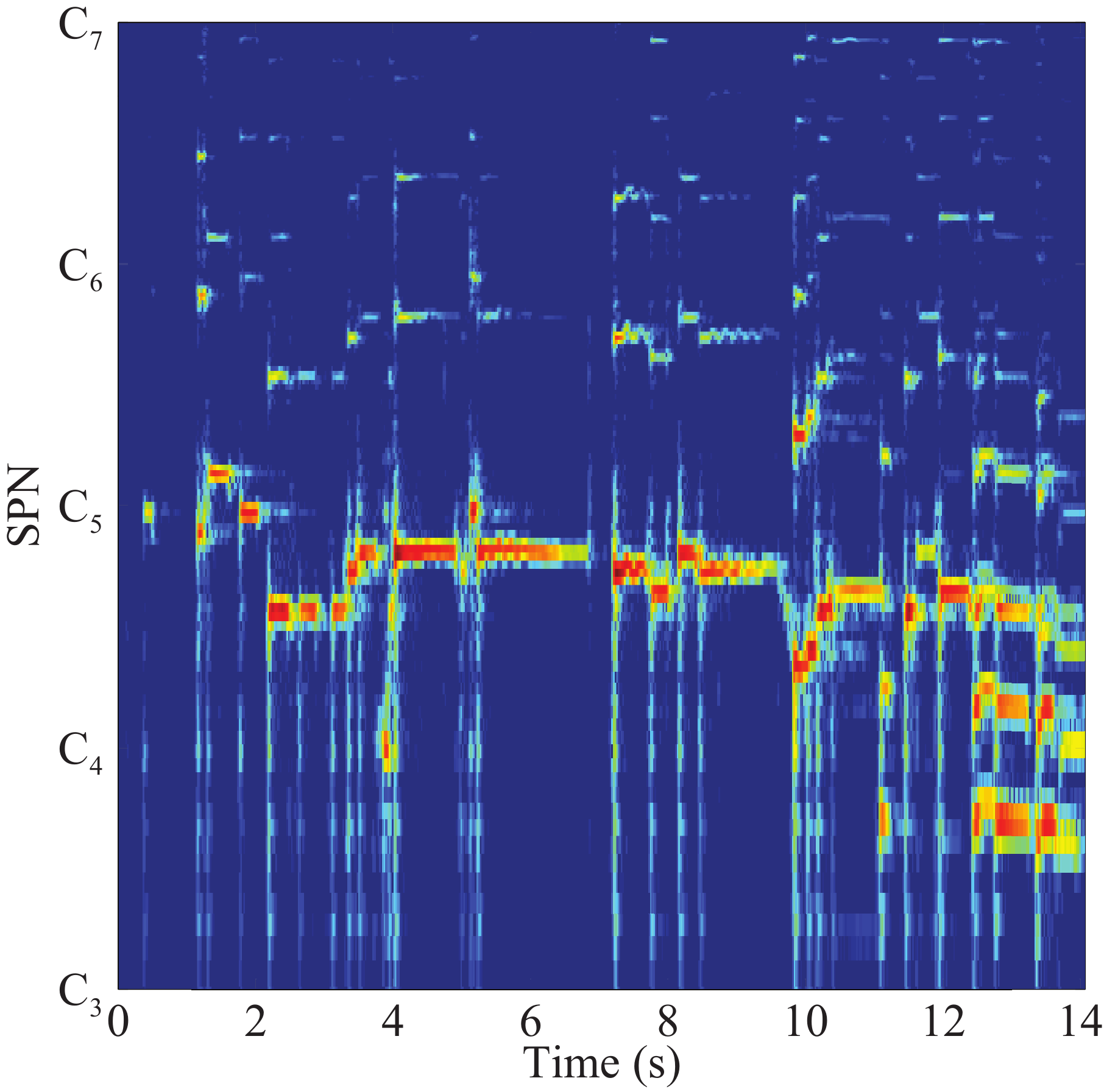}\label{fig:pg_spectrogram}}
\qquad\qquad
\subfigure[Manual annotation (binary mask)]{
	\includegraphics[height=.35\textwidth]{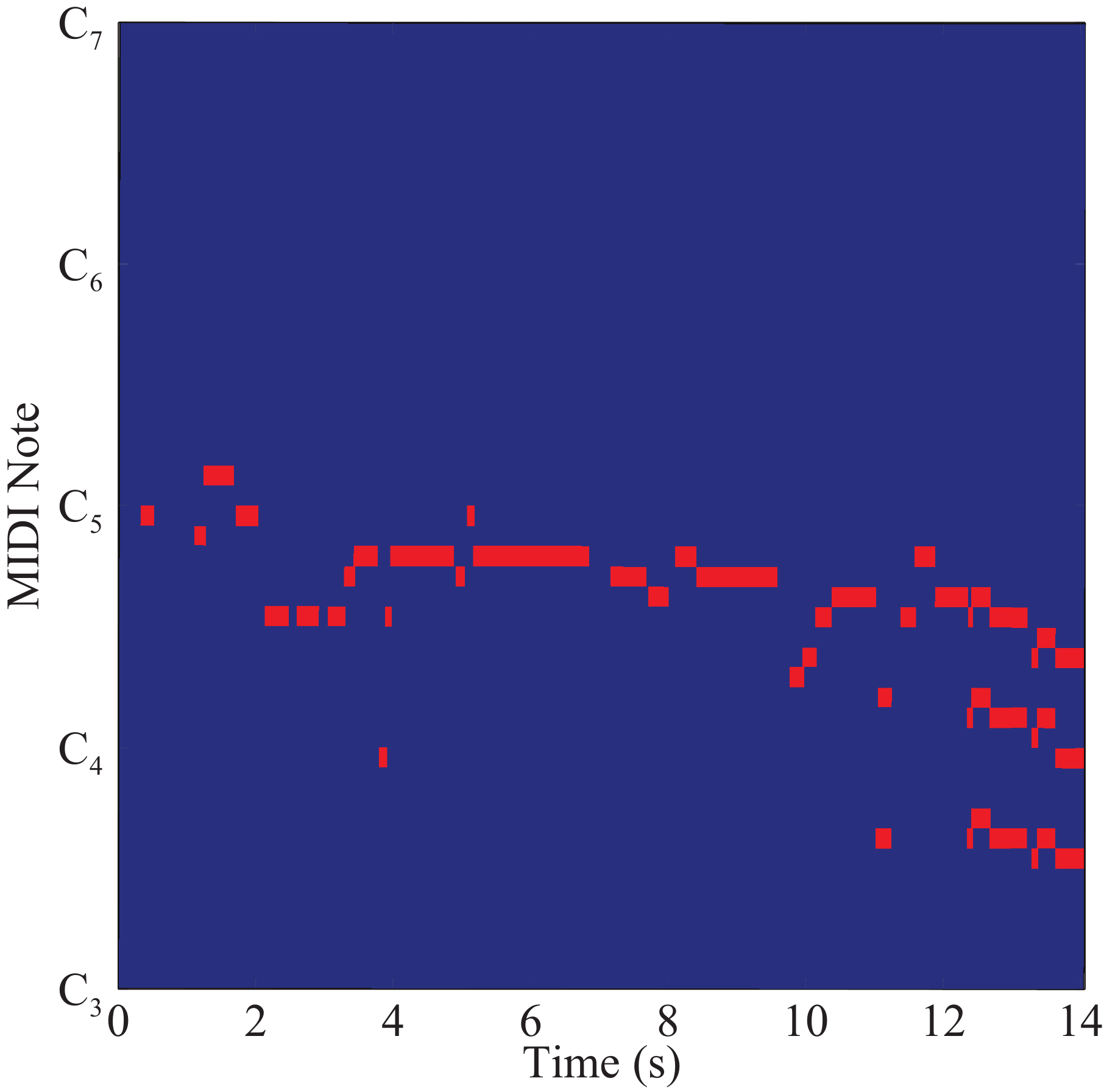}\label{fig:pg_mask}}
\\
\subfigure[Power-weighted sinc (logarithmic)]{
	\includegraphics[height=.35\textwidth]{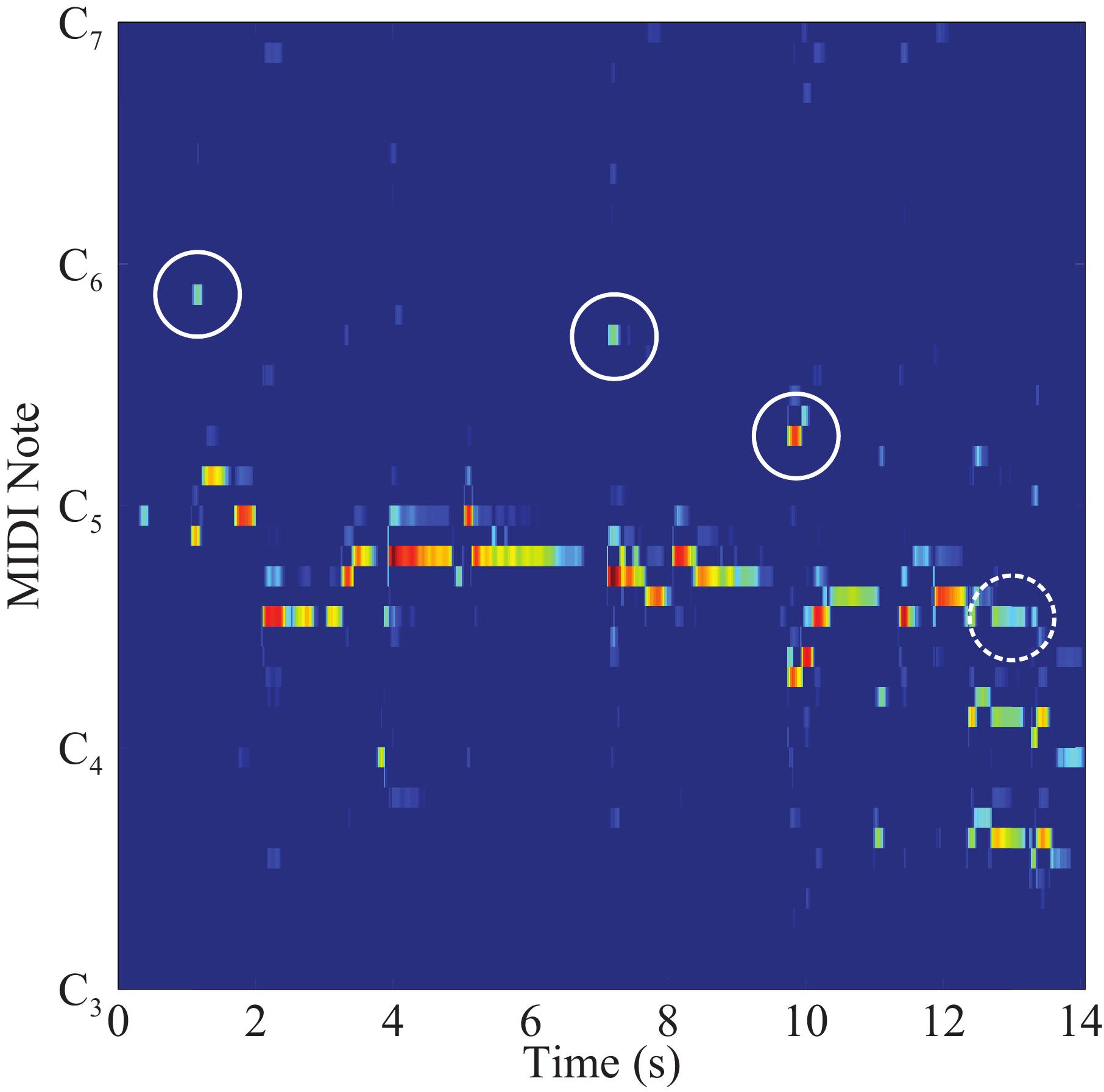}\label{fig:pg_sincw}}
\qquad\qquad
\subfigure[Power-weighted bident (logarithmic)]{
	\includegraphics[height=.35\textwidth]{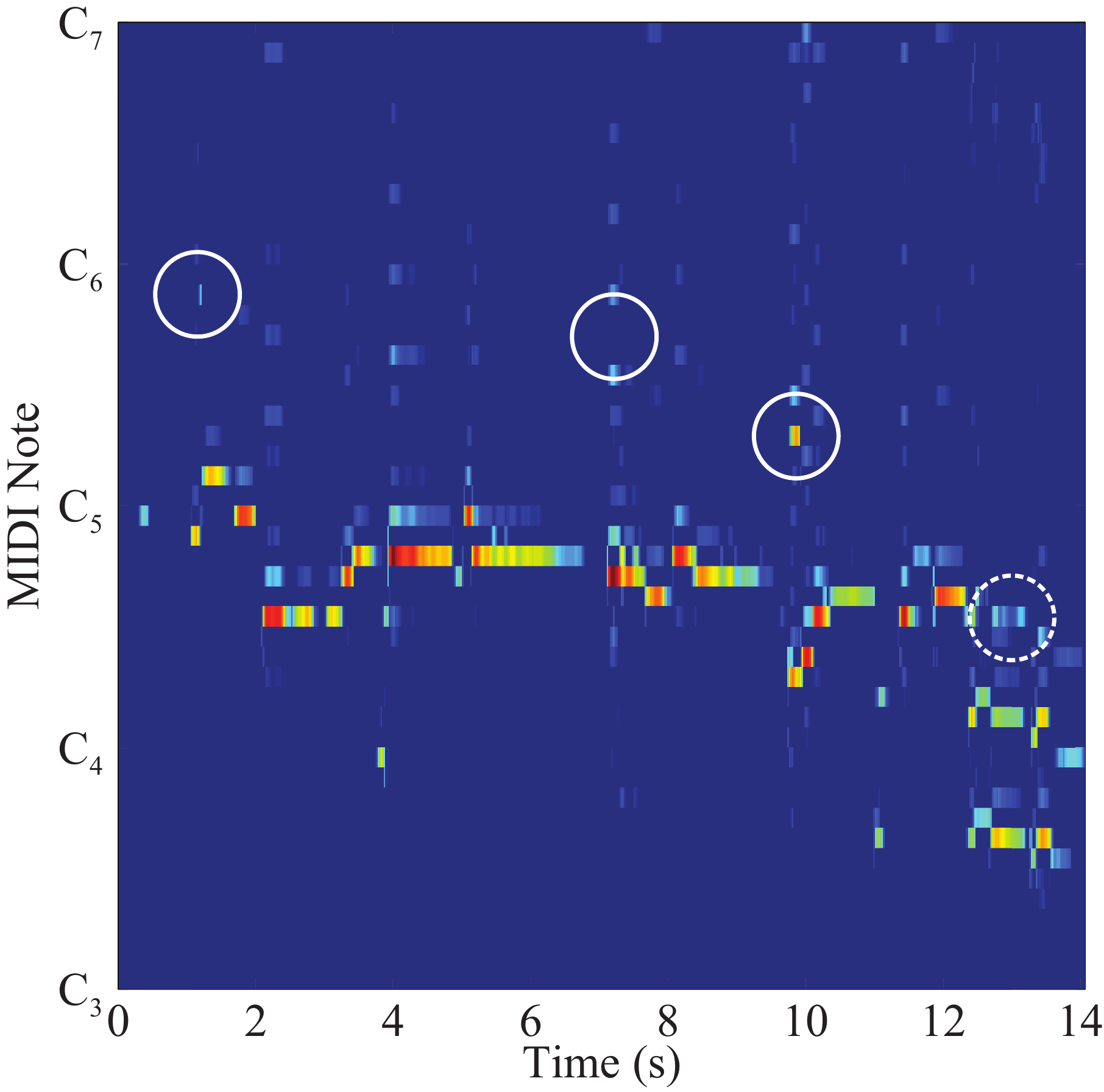}\label{fig:pg_bidentw}}
\\
\subfigure[Power-invariant sinc (linear)]{
	\includegraphics[height=.35\textwidth]{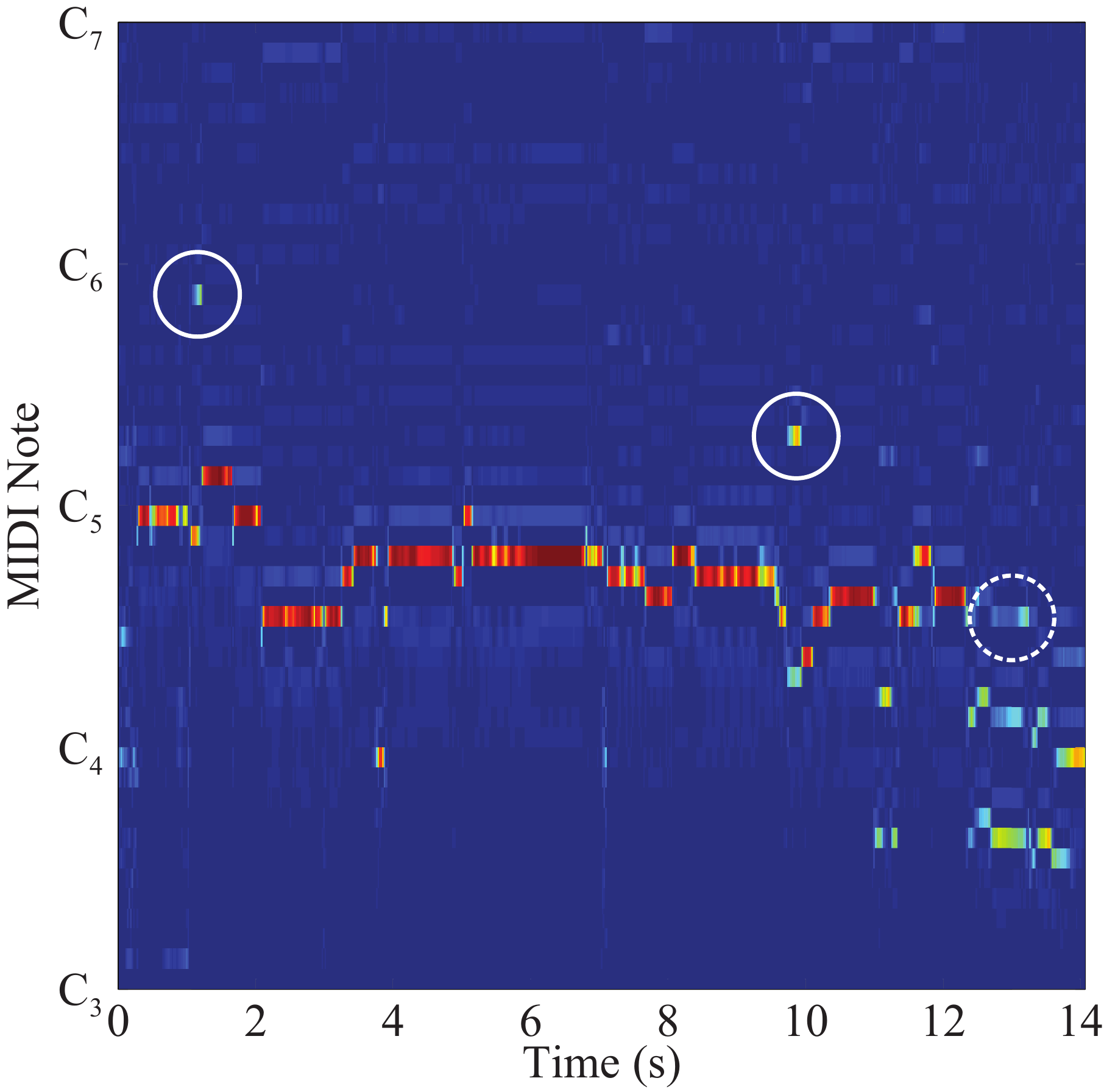}\label{fig:pg_sinc}}
\qquad\qquad
\subfigure[Power-invariant bident (linear)]{
	\includegraphics[height=.35\textwidth]{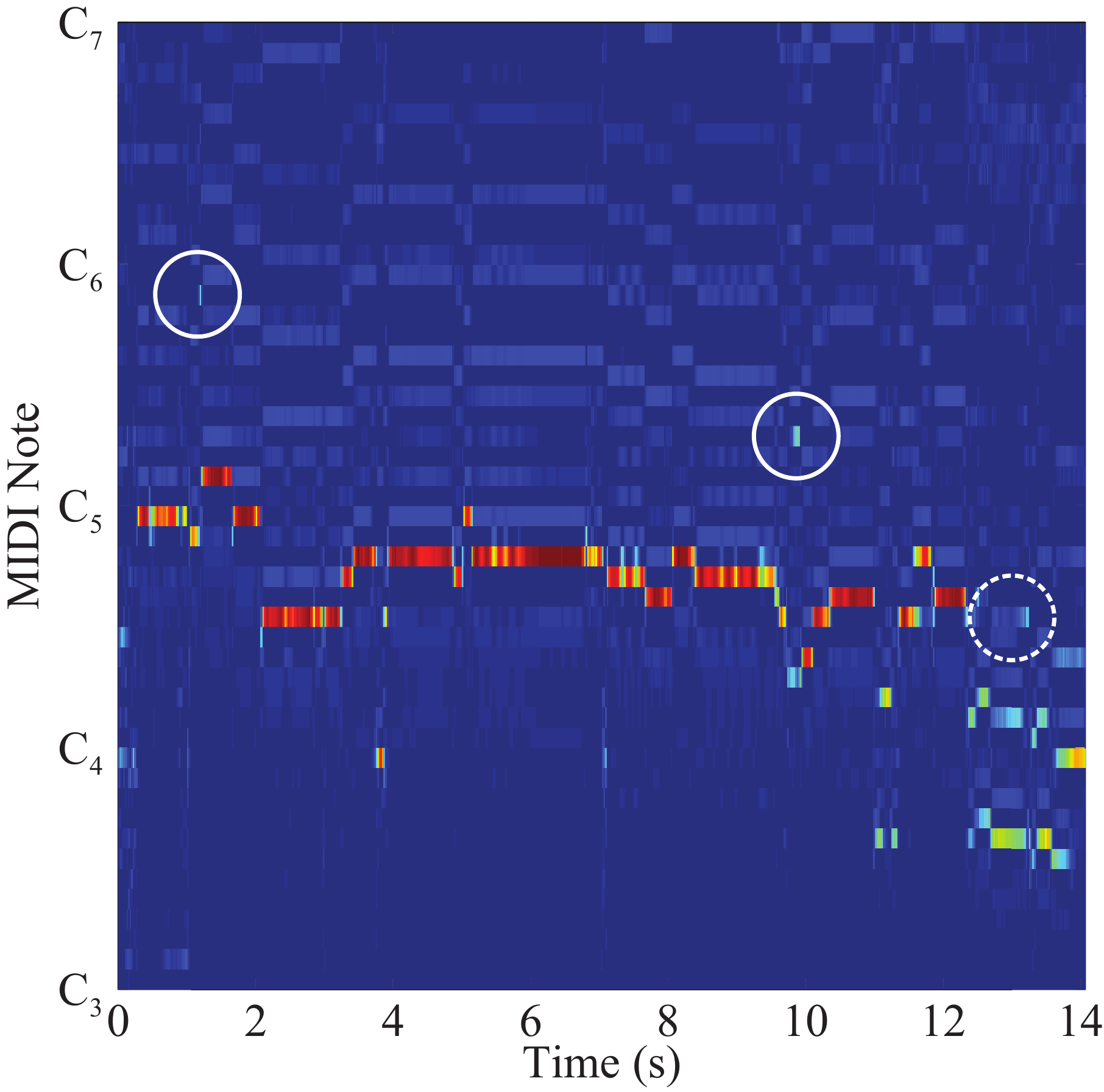}\label{fig:pg_bident}}
\caption{Results from the pitch analysis of a jazz guitar solo (``Solar'').}
\label{fig:pitchgram}
\end{figure*}

Fig.~\ref{fig:pitchgram} shows the results from the pitch analysis of a jazz guitar solo. In Fig.~\ref{fig:pg_spectrogram} we see the signal's $\log$-spectrogram. The ideal binary mask\footnote{\textcolor{red}{The binary mask was obtained by fine tuning an automatic transcription by hand and ear in a digital audio workstation.}} is shown in Fig.~\ref{fig:pg_mask}. Figs.~\ref{fig:pg_sincw}--\ref{fig:pg_bident} illustrate the different variants of the pitchgram. The left column shows the results obtained with the $\sinc$ filter, while the right column shows the results obtained with the bident filter, respectively. In the middle row, both the pitchgrams were computed based on a harmonically weighted ACF. In the lower row, the ACF is normalized by the mean signal power. Furthermore, the ACF was compressed in all four cases. The spectrogram was computed using a 4096-point discrete Fourier transform and a Hamming window of equal length with an overlap of 75 \% between consecutive blocks. The pitchgram uses the same time resolution. 

Looking at the pitchgram, one can easily identify single notes, arpeggios, and chords. At various time instants, the pitch is modulating between the unison and the minor second (one semitone)---a clear indication of vibrato, see Fig.~\ref{fig:pg_spectrogram}. For the most part, this translates to amplitude variations in the pitchgram. The issue with vibratos is how not to interpret (periodic) variations in amplitude as a tremolo, which would otherwise make a single note appear as several notes in the transcription. As a similar case, hammer-ons, which may be not strong enough to be recognized as note onsets, and pull-offs, which may not make the tone decay to a low power level before the string is struck again, both require special attention. This brings up the question whether, to a certain degree, the affected notes could be associated with a playing technique, such as vibrato, legato, etc.\ If this was the case, one could then carry out a more sophisticated performance analysis with regards to articulation and expression, something that is beyond pitch and onset. One may also try to identify technical errors, which otherwise would result in phantom notes. The latter are provoked by fast finger moving and sloppy finger placement when plucking. It can also be seen that sliding chords (glissando) are more tricky to handle, because the note onsets are slurred and the fingers encroach on neighboring tones. 

The bident successfully eliminates, or at least alleviates, the octave ambiguities that are due to an accentuated (first) overtone, see white circles. Occasionally, this also happens at the expense of the fundamental, which might perish together with the overtones; the dashed circle points to such a case where the upper forth disappears almost completely. In part, this type of effect is due to the chosen signal representation: the ACF. An overtone can come out stronger (in the spectrogram) than it is actually heard, when the phase relations between the harmonics are impure. Certainly, we can also discover some artifacts, which are a consequence of the side lobes of the $\sinc$ filter and the resulting ripple of the bident. As a final remark, we would like to mention that note velocity is useful for the assessment of dynamics in a performance or for the detection of a particular form of accentuation.

\color{red}
\subsection{Comparison with NMF}

\begin{figure}[!t]
\centering
\includegraphics[width=.7\columnwidth]{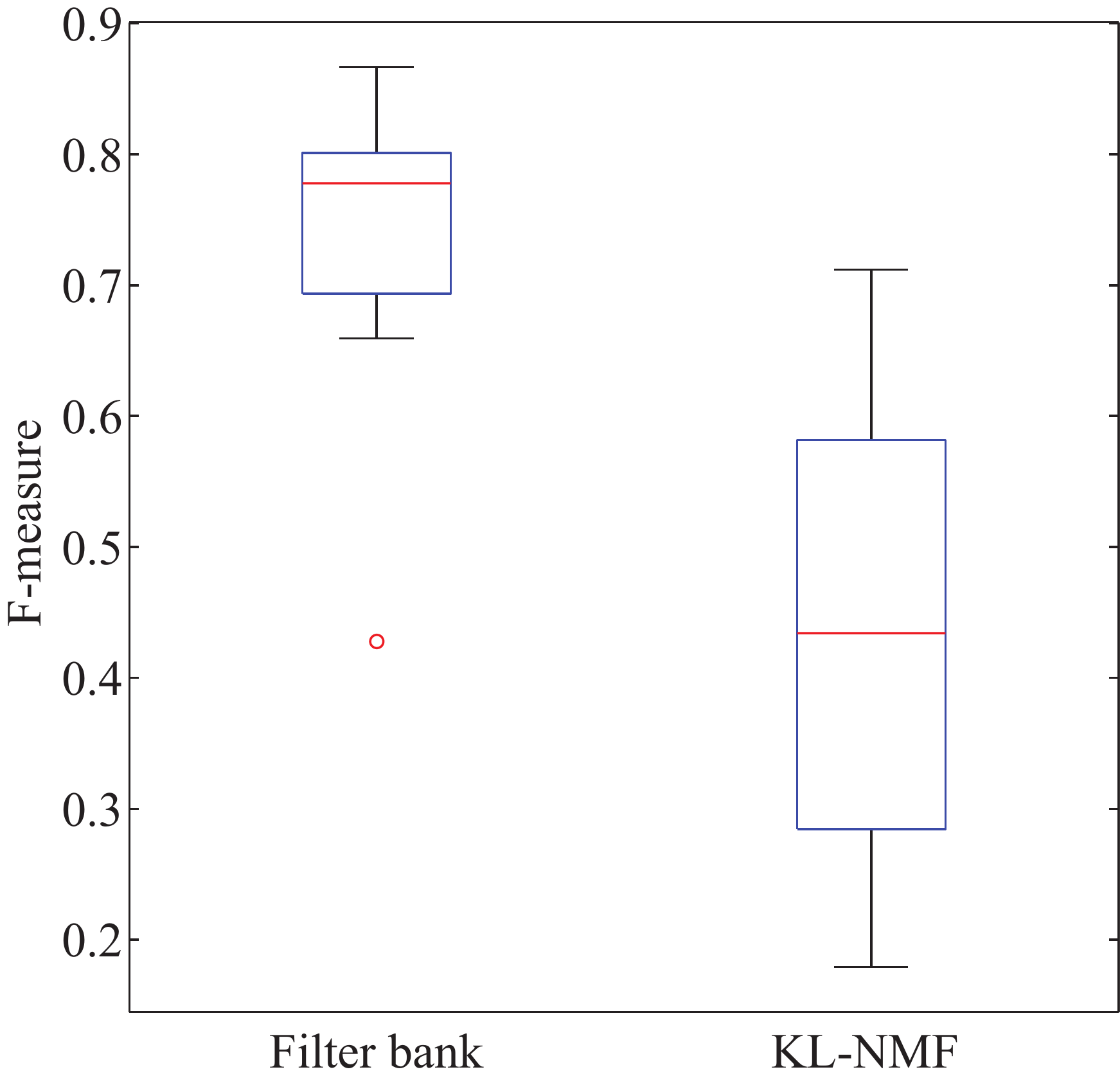}
\caption{\textcolor{red}{Pitch detection accuracy: comparison between the filter bank and the KL-NMF with multiplicative updates w.r.t.\ the F-measure.}}
\label{fig:cmp_nmf}
\end{figure}

To quantify the improvement that one can achieve with the filter bank approach in comparison to an established tool for (polyphonic) music transcription, such as the NMF, we carry out the following experiment. Firstly, we compute the power-weighted and the power-invariant pitchgram, as in Figs.~\ref{fig:pg_bidentw} and \ref{fig:pg_bident}, for the corpus we resort to later in our performance evaluation, see Section~\ref{sec:corpus}. The two representations are then multiplied by each other and the result is normalized by the Frobenius norm. In a way, this is equivalent to computing the posterior pitchgram (the power-invariant pitchgram may also be interpreted as the tonality likelihood). The new pitchgram is then smoothed, see Section~\ref{sec:pitch_detection}, and finally compared to a threshold with a binary decision, to yield a binary mask. It should be noted that the power-weighted pitchgram accounts for the duration of notes. Secondly, the NMF is computed. A standard procedure is to take the magnitude spectrum for the time-frequency representation (TFR) and to use the Kullback--Leibler divergence as distance metric for the factorization.  In this regard, we use Lee and Seung's (original) algorithm with multiplicative update rules from the NMFlib.\footnote{\color{red}\url{http://www.ee.columbia.edu/~grindlay/code.html}} The algorithm is run ten times (with different initializations) for each signal and the factorization with the smallest error is kept. The rank of the factorization is set equal to the number of notes in the signal, which is obtained from the reference transcription. In an intermediate step, the activation matrix is converted into a time-pitch representation and then post-processed in the exact same manner as the pitchgram. The frequency resolution for the NMF is 10.8 Hz per bin and the time resolution for both the pitchgram and the NMF is 1024 samples (23.2 ms). The outcome of the comparison is shown in Fig.~\ref{fig:cmp_nmf}. On each box, the central mark is the median, and the edges of the box are the 25th and 75th percentiles. The whiskers indicate the most extreme data points, while outliers are plotted as circles. The F-measure is defined in Section~\ref{sec:metrics}. 

Quite clearly, the posterior pitchgram is way more reliable than the NMF under the employed protocol. The result can be explained as follows. Firstly, the error of the NMF is defined as the (statistical) distance between the original spectrogram and its factorization---not on the factors. Given that the rank, i.e.\ the number of factors, was set equal to the ideal number of perceptually distinct notes, pitch modulations within half a semitone can easily result in wrong pitch. This is because the spectral note profile of a neighboring tone can be closer to a pitch-bended version of the original tone, or simply because the spectral imprint of a similar note gives a smaller error in reference to the original spectrogram. Secondly, it is common knowledge that the attack (transient) and sustain (tonal) parts of a note should be represented with different bases. If only the tonal part can be assigned to a basis, note onsets become another source of error. Finally, it should be noted that better results can be obtained, if the NMF bases are pre-learned on isolated note spectrograms with different accentuations and/or playing techniques. The advantages of the pitchgram are:
\begin{itemize}
\item Number of notes (rank) not needed as prior knowledge;
\item No ambiguity in note profiles (dictionary);
\item No convergence issues (convexity/initialization);
\item No distinction between attack and sustain necessary.
\end{itemize}

\color{black}

\section{Automatic Transcription}
\label{sec:transcription}

In this section, we \textcolor{blue}{provide some} technical details on how to realize an \textcolor{blue}{\emph{exemplary}} transcriber that resorts to the pitchgram. \textcolor{red}{To counter the issue of signal level dependency of the tuning parameters, we suggest to normalize the input signal level to a reference signal level and to perform the detection steps on both the power-invariant and the power-weighted pitchgrams, whenever ambiguities arise.} 
The complexity of the transcriber and its robustness is largely determined by the onset and decay detection blocks. 


\subsection{Pitch Detection}
\label{sec:pitch_detection}

The detection of a note should be reliable, and so it requires a robust decision criterion. With this goal in mind, we define the following normalized derivative:
\begin{equation}
\dot{Y}(m, p) = \frac{\Delta Y(m, p)}{\overline{Y}(m, p)} = \frac{Y(m, p) - Y(m - 1, p)}{\color{blue}{\left[G(\cdot) \ast Y(\cdot, p)\right]} {\left(m\right)}} \text{,} 
\label{eq:derivative}
\end{equation}
where $\Delta Y(m, p)$ is the first-order derivative of $Y$ and $\overline{Y}(m, p)$ is the smoothed value of $Y$ in $(m, p)$. $G$ is the smoother and $\ast$ denotes linear convolution, respectively. One can also use the median filter in place of the linear smoother. A note onset is subject to the following conditional \textcolor{blue}{(if-then)} statement: 
\begin{equation}
Y(m, p) > \epsilon_1 \, \wedge \, \dot{Y}(m, p) > \epsilon_2 \rightarrow \mathrm{onset}(m, p)
\label{eq:onset}
\end{equation}
with $\epsilon_1$ and $\epsilon_2$ being empirical threshold values. So, an onset is defined by a sufficiently high pitch score and a steep slope. In addition, one may also check whether the slope is rising. 

\subsubsection{Transient Noise}

In most cases, the onset of a note is accompanied by transient noise, which spreads over the neighboring pitches. This is especially the case for the pitchgram that was obtained from the spectrogram. To make sure that transient noise is not mistaken for a note onset, we check whether the following condition is fulfilled:
\begin{equation}
\Delta Y(m, p) > \Delta Y(m, p - 1) \, \wedge \, \Delta Y(m, p) > \Delta Y(m, p + 1) 
\label{eq:transient}
\end{equation}
subject to the existence of $p - 1$ and $p + 1$. By transient noise we mean the score spread that appears at the onset of plucked or picked notes. 

\subsubsection{Fretting Noise}

Another form of noise that can falsify the detection of a note onset is fretting noise. Fretting noise appears when the finger is moved along the guitar neck and the string is pressed down behind a fret. It is also typical for hammer-ons and pull-offs. As the origins of fretting noise are different from the origins of transient noise, we introduce another condition to make sure that fretting noise is not mistaken for pitch, which is
\begin{equation}
Y(m, p) > Y(m, p - 1) \, \wedge \, Y(m, p) > Y(m, p + 1) 
\label{eq:fretting} 
\end{equation}
subject to the existence of $p - 1$ and $p + 1$. Just as \eqref{eq:transient}, \eqref{eq:fretting} implicitly assumes that the tonality of noise is lower than the tonality of the actual pitch. 

\subsubsection{Octave Errors}

A simple method to avoid octave errors \textcolor{red}{in monophonic music material} is to ensure that there was no note already detected twelve semitones or one octave below the current pitch value $p$. In that case, the new onset should be discarded. 

\subsubsection{Note Velocity}

According to the MIDI standard, velocity is a measure of how rapidly and forcefully a key on a keyboard is pressed when it  is idle. In the same way, we may describe with which force a string was hit at note onset\textcolor{blue}{:} 
\begin{equation}
\mathrm{onset}(m, p) \rightarrow \mathrm{velocity}(m, p) \propto Y(m, p) \text{,} 
\label{eq:velocity}
\end{equation}
where $Y(m, p)$ is the power-weighted pitchgram from \eqref{eq:pitchgram}. \textcolor{red}{In \eqref{eq:velocity}, it is assumed that the velocity (force) is proportional to the measured power at onset. Another way to obtain it, is by integrating the energy over the entire note duration.} 

\subsection{Pitch Tracking}

After detection of an onset, the respective note is deemed to be active, i.e.\ key pressed or string vibrating, until its energy falls below a certain threshold value and the first derivative is sufficiently large and negative, see \eqref{eq:derivative}. And so, the condition for a note decay is as 
\begin{equation}
Y(m, p) < \epsilon_3 \, \wedge \, \dot{Y}(m, p) < \epsilon_4 \rightarrow \mathrm{decay}(m, p) \text{,} 
\label{eq:decay}
\end{equation}
where $\epsilon_3$ and $\epsilon_4$ are again empirical threshold values. It may further be wise to check whether the slope in falling to avoid ambiguities. Accordingly, the note duration is given by 
\begin{equation}
d(p) = m_\mathrm{off} - m_\mathrm{on} \text{,} 
\end{equation}
where $m_\mathrm{on}$ is the onset instant and $m_\mathrm{off}$ the decay instant. 

\subsubsection{Vibrato}

On various occasions, the player may make use of vibrato as a means of musical expression. Depending on the hop size of the sliding DFT, i.e.\ the time resolution of the pitchgram, and the rate of vibrato, the pitch may appear and disappear with a certain regularity. Given \eqref{eq:decay} only, vibrato will be transcribed almost surely as tremolo, i.e.\ a quick repetition of the same note. Since the typical extent of vibrato is less than a semitone either side of the note, we should avoid mistaking vibrato for tremolo by replacing the first hypothesis in \eqref{eq:decay} by 
\begin{equation}
Y(m, p) + Y_\mathrm{vib}(m, p) < \epsilon_3 \text{,} 
\end{equation}
where
\begin{equation}
Y_\mathrm{vib}(m, p) = \max{\left\{Y(m, p - 1), Y(m, p + 1)\right\}} 
\end{equation}
on the supposition that there are no active notes one semitone above and below the current pitch value $p$.

\subsubsection{Pruning}

Once we made the decision that a note is no longer active, we might verify that is not too short, and hence not an artifact of an imperfect time-pitch representation. We call it pruning in reference to machine learning terminology. The condition for a note to be valid is simply 
\begin{equation}
d(p) > d_{\min} \text{.} 
\end{equation}
It should be noted that if $d_{\min}$ is too long, notes belonging to an arpeggiated chord might get lost as well. 

\subsection{Transient Detection}
\label{sec:transient}

The condition in \eqref{eq:onset} for the note onset may not always be sufficient. This is usually the case when one and the same note is rapidly reiterated (tremolo) or, in more general terms, when the string is picked or plucked again before it stopped vibrating at a high amplitude. In such a case, the slope of $Y$ in point $(m, p)$ is not steep enough for the second hypothesis in the conditional statement to be true. Then, it is helpful to compute a transient score by marginalizing out the pitch from $Y(m, p)$, 
\begin{equation}
Y(m) = \sum\nolimits_p{Y(m, p)} \qquad \forall \: p \text{,} 
\end{equation}
and by evaluating the normalized derivative of the marginal score $\dot{Y}(m)$ in addition to \eqref{eq:onset}. 
\textcolor{blue}{It must be admitted, however, that an additional transient detector might still be necessary.} 

\section{Performance Evaluation}
\label{sec:evaluation}

In this section, we present the results of the performance evaluation of our rule-based transcriber in regard to another state-of-the-art transcriber for the guitar. Given the complexity and the number of processing steps, or layers, of most of the melody detection algorithms, we would consider it rather unfair to make a comparison between those systems and our transcriber, which operates \textcolor{blue}{\emph{exclusively}} on the pitchgram. In addition, we would like to compare our transcriber to a system that was \textcolor{blue}{\emph{tuned to the guitar}}, and not the piano, and which does not require prior training. \textcolor{red}{We also include PYIN.} 

\subsection{Algorithms}

In our evaluation, we \textcolor{blue}{compare} the output of the following transcription systems: 
\begin{description}
\item[A] The reference transcriber \cite{Ozaslan2011,Serra2013}; 
\item[B] Our (decision-based) transcriber with the \textcolor{blue}{pitchgram being computed using the fast Fourier transform}; 
\item[C] The same transcriber as in \emph{B}, but with the \textcolor{blue}{original time}-domain pitchgram as input \textcolor{blue}{(see Section~\ref{sec:filter_bank})}; 
\color{red}
\item[D] PYIN \cite{Mauch2014}, which is an improved version of YIN. 
\end{description}
The reference transcriber is an updated version of the extraction module used in the two references. It was tested with both \textcolor{blue}{\emph{nylon and steel}} guitar strings. The extraction module combines the outputs of an onset detector and a pitch detector. The pitch detector is YIN. The pitch trajectory in between two note onsets is mapped to a single pitch value in order to obtain a piano-roll-like representation of the input signal. The onset detector is described in \cite{Duxbury2003}. It should be noted that the algorithm is of great value for the comparison, because 1) YIN is the de-facto standard in most monophonic transcription tasks and 2) given the low degree of polyphony of our test material, it is expected to perform \textcolor{blue}{well}. The same can be said about the onset detector. \textcolor{red}{PYIN is a more recent variant of the YIN algorithm. It was invoked in the form of a Vamp plugin\footnote{\color{red}\url{https://code.soundsoftware.ac.uk/projects/pyin}} from within the Sonic Visualiser\footnote{\color{red}\url{http://www.sonicvisualiser.org/}} environment, which offers the possibility to export the transcription to a MIDI file. The main difference between YIN and PYIN is that the latter has a prior for the threshold instead of a fixed value. We chose a mean of 0.15 for the beta prior, as suggested by the authors. }

\subsection{Corpus}
\label{sec:corpus}

As our goal was to develop a transcriber that captures the subtle nuances of a jazz guitar in the first place, we have created a new data corpus consisting of ten jazz guitar excerpts (phrases) from solos that were performed and recorded at Sony CSL in Paris \cite{Gorlow2016_JazzGuitarLicks}. The excerpts were annotated manually, so as to have a reference transcription. The note onsets were determined according to their perceptual salience. The same applies to the duration of single notes. The data contains both monophonic and polyphonic (chordal) melody lines, although the number of chordal passages is small. The manual transcription, which is arguably not perfect, serves as the ground truth in our experiment. The data corpus, the manual annotation, and the corresponding output from the algorithms mentioned above are \textcolor{red}{all} part of the electronic appendix to the \textcolor{blue}{manuscript}. \textcolor{red}{The reference transcriber and our own transcriber were tuned by hand to perform well over the entire data set.} 

\subsection{Metrics}
\label{sec:metrics}

The algorithms were compared in terms of the following performance metrics. Prior to computing the scores, the provided MIDI files were converted to binary masks. The time axis was quantized. 
\begin{description}
\item[F-measure] In binary classification, the F-measure indicates the accuracy of a system under test. It is defined as the harmonic mean of precision and recall: 
\begin{equation}
F = 2 \cdot \frac{\mathrm{precision} \cdot \mathrm{recall}}{\mathrm{precision} + \mathrm{recall}} 
\end{equation}
with
\begin{equation}
\begin{aligned}
&\mathrm{precision} \\
&\quad {} = \frac{\abs{{\left\{\mathrm{detected\ pitches}\right\}} \cap {\left\{\mathrm{reference\ pitches}\right\}}}}{\abs{\left\{\mathrm{detected\ pitches}\right\}}}
\end{aligned}
\label{eq:precision}
\end{equation}
and
\begin{equation}
\begin{aligned}
&\mathrm{recall} \\
&\quad {} = \frac{\abs{{\left\{\mathrm{detected\ pitches}\right\}} \cap {\left\{\mathrm{reference\ pitches}\right\}}}}{\abs{\left\{\mathrm{reference\ pitches}\right\}}} \text{.} 
\end{aligned}
\end{equation}
Precision is the fraction of detected pitches that are in the reference transcription, i.e.\ which are relevant, while recall is the fraction of relevant pitches that are detected. Accordingly, precision is often interpreted as a measure of exactness or \emph{quality} as opposed to recall, which is seen as a measure of completeness of \emph{quantity}. In our case, a detected pitch is a point with a value of {1} in the time-pitch plane, which is equivalent to a piano-roll representation. Undetected pitches then again are points with a value of {0}. The F-measure attains its best value  at one and its worst value at zero. To account for the relevance of the detected pitches, we also weight each detected pitch by the MIDI velocity of the corresponding note during the computation of \eqref{eq:precision}\textcolor{red}{, where the velocity is given by \eqref{eq:velocity}. By doing so, we make sure that softer and more subtle notes, which are more likely to go unheard, have lesser impact on the precision.} 
\item[Error score] The error score is a common metric in speech-to-text transcription \cite{RT2009}. It is nothing else but the word error rate that has established itself as the ultimate accuracy metric in the field of automatic speech recognition. Respectively, the error score is computed as: 
\begin{equation}
\begin{aligned}
E &= \frac{\abs{\{\mathrm{pitch\ substitutions}\}} + \abs{\{\mathrm{pitch\ deletions}\}}}{\abs{\{\mathrm{reference\ pitches}\}}} \\
&\qquad {} + \frac{\abs{\{\mathrm{pitch\ insertions}\}}}{\abs{\{\mathrm{reference\ pitches}\}}} \text{.} 
\end{aligned}
\end{equation}
Here again, we assess the accuracy of the algorithms based on their binary masks. Substitutions are pitch errors, such as octave or semitone errors, deletions are missing pitches, and insertions are false positives, i.e.\ pitches that should not be. The latter appear, e.g., when a note is detected before the actual onset, or when a note's actual duration is exceeded. Unlike the F-measure, the error score can be larger than one. 
\end{description}

\subsection{Results} 

\begin{figure*}[!t]
\subfiguretopcaptrue
\centering
\subfigure[Pitch detection]{
	\includegraphics[height=.35\textwidth]{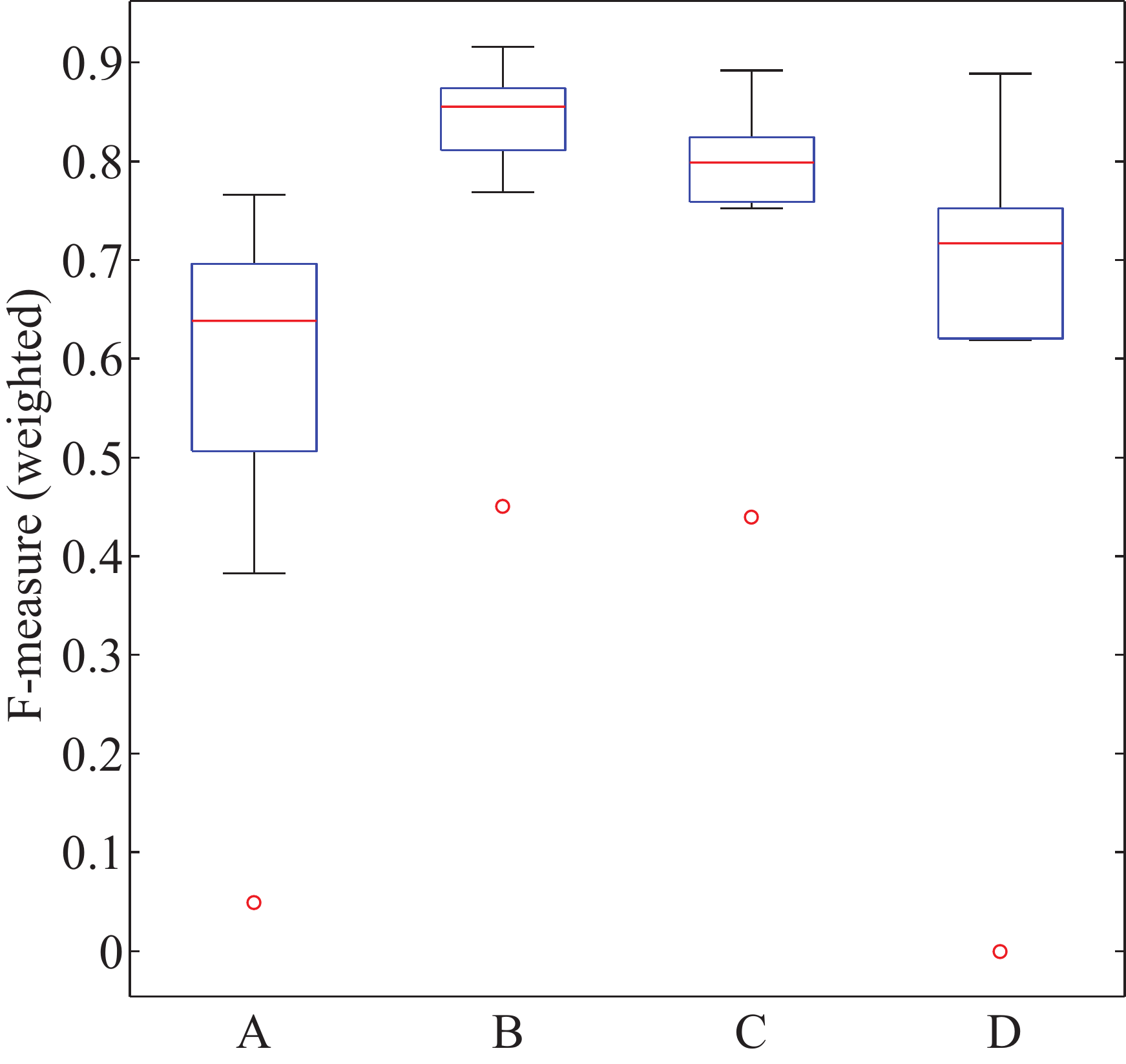}
	\qquad\qquad
	\includegraphics[height=.355\textwidth]{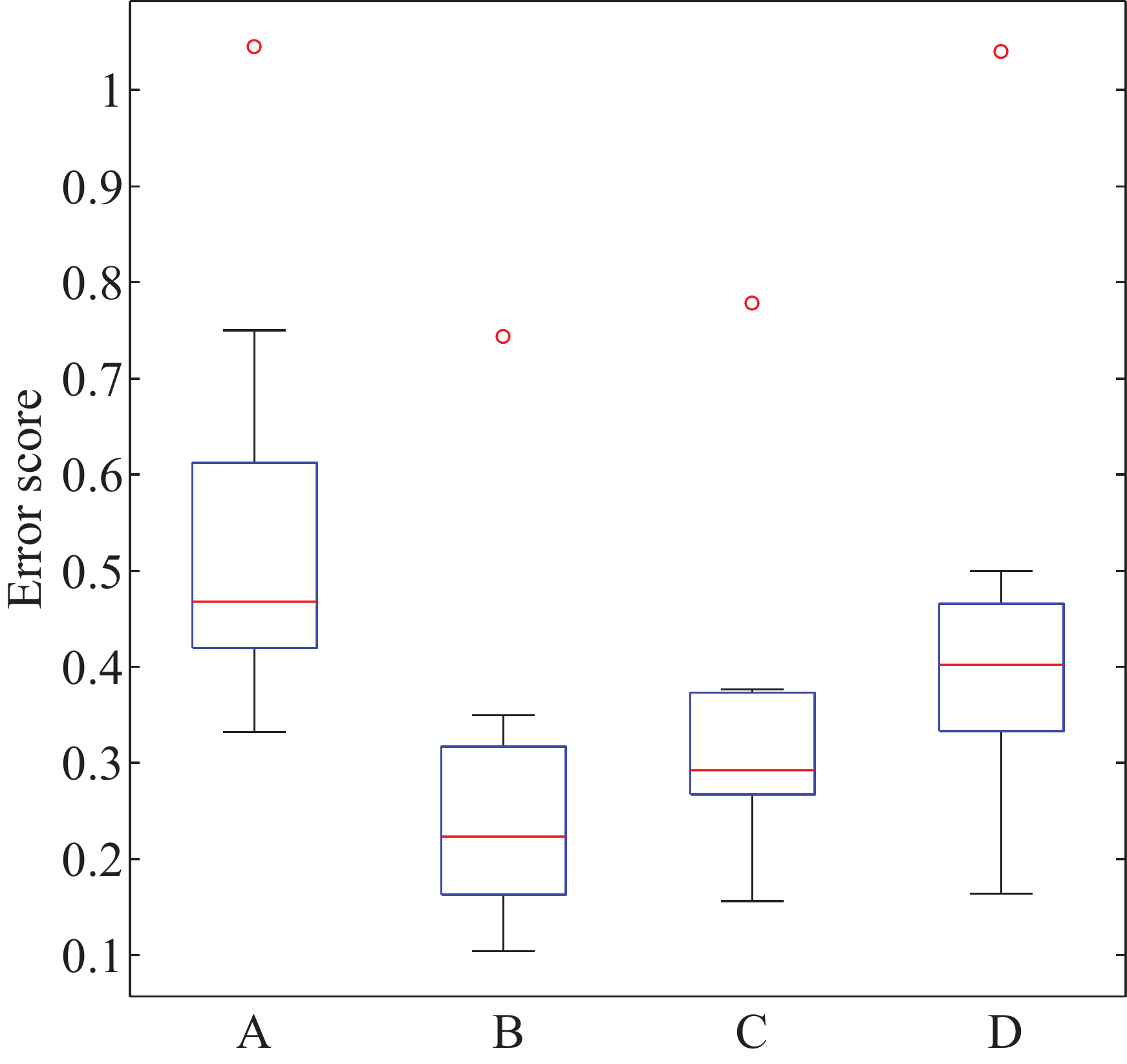}
	\label{fig:pitch}}
\\
\subfigure[Onset detection]{
	\includegraphics[height=.35\textwidth]{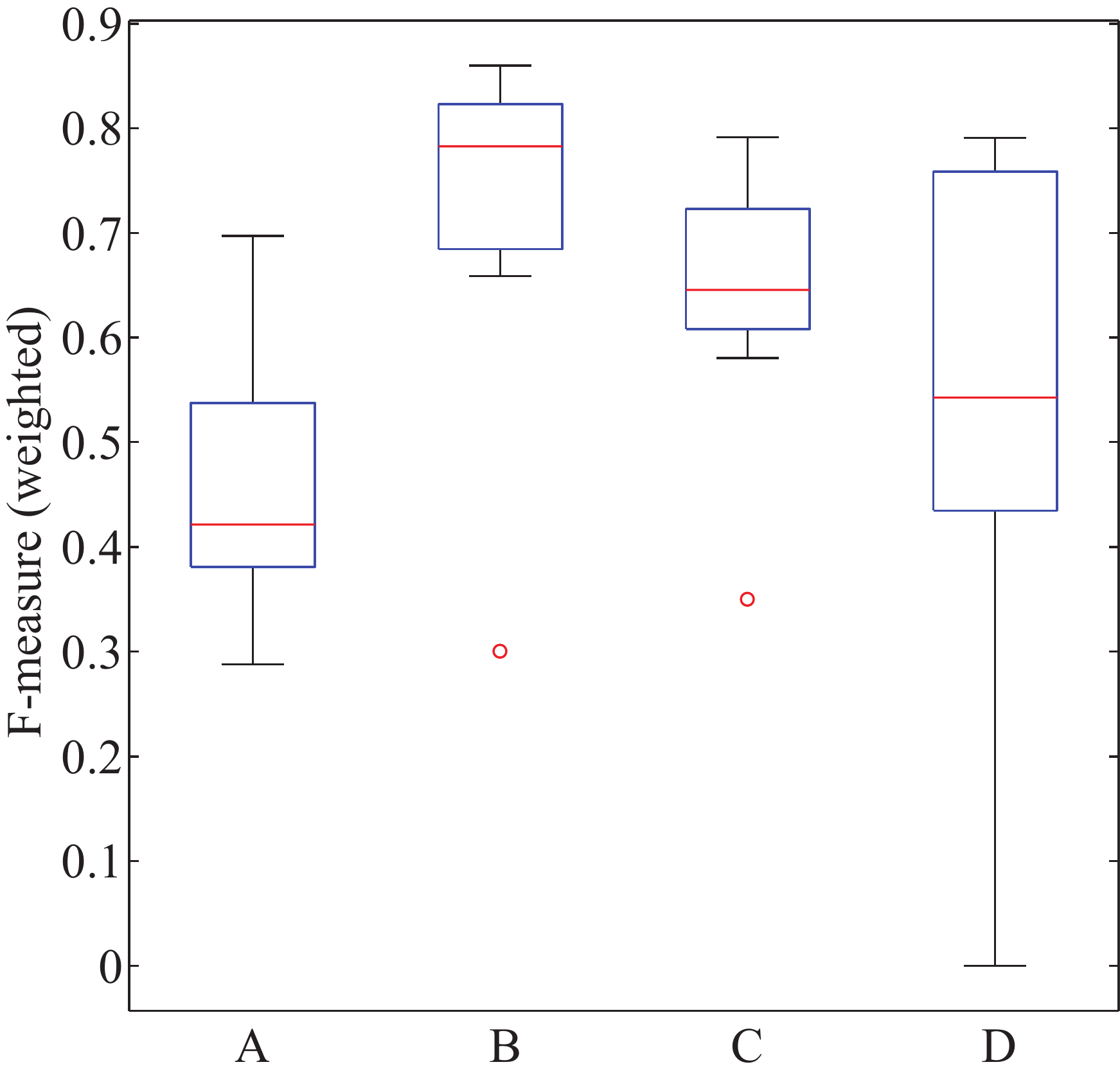}
	\qquad\qquad
	\includegraphics[height=.35\textwidth]{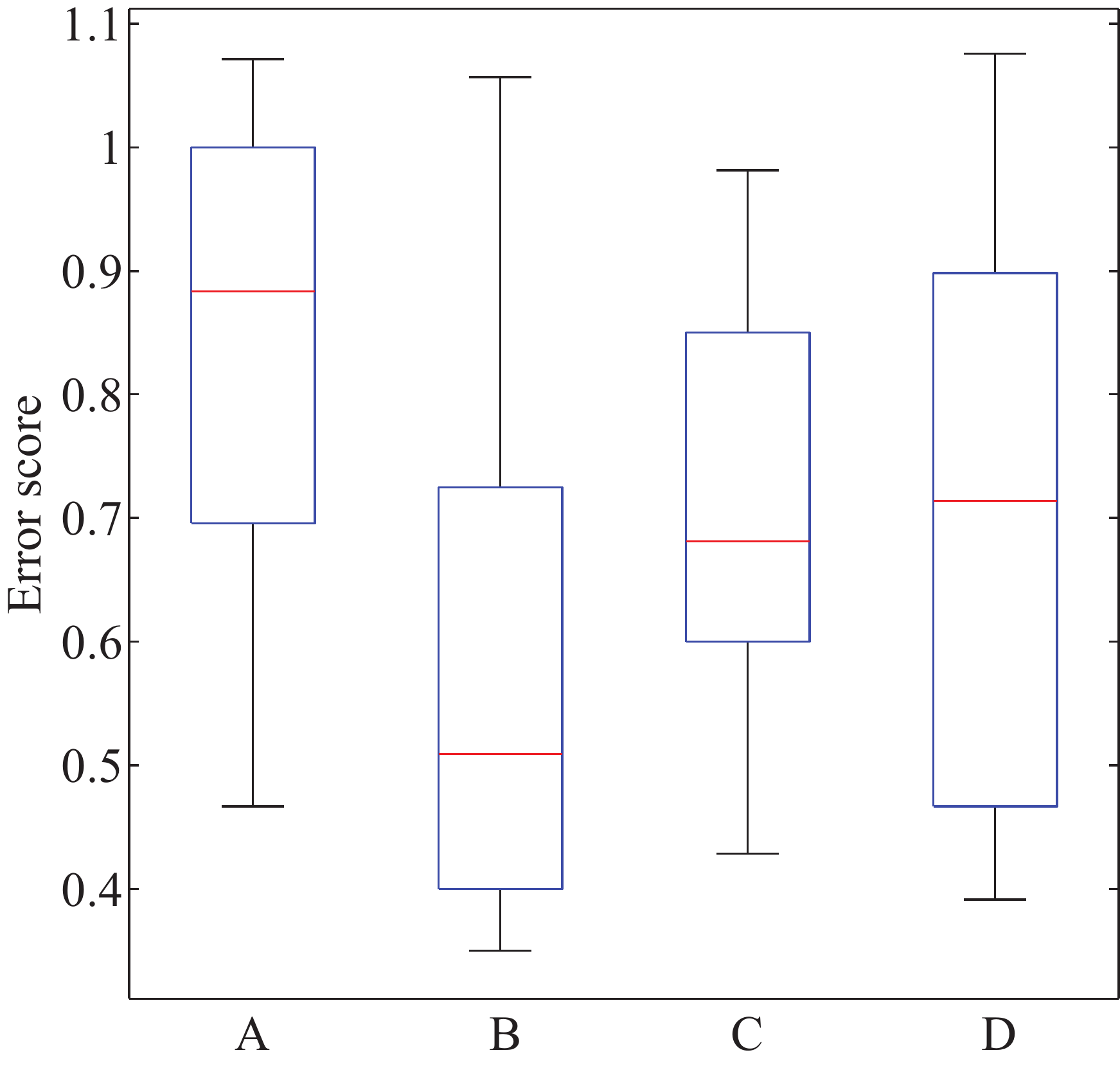}
	\label{fig:onset}}
\\
\subfigure[Decay detection]{
	\includegraphics[height=.35\textwidth]{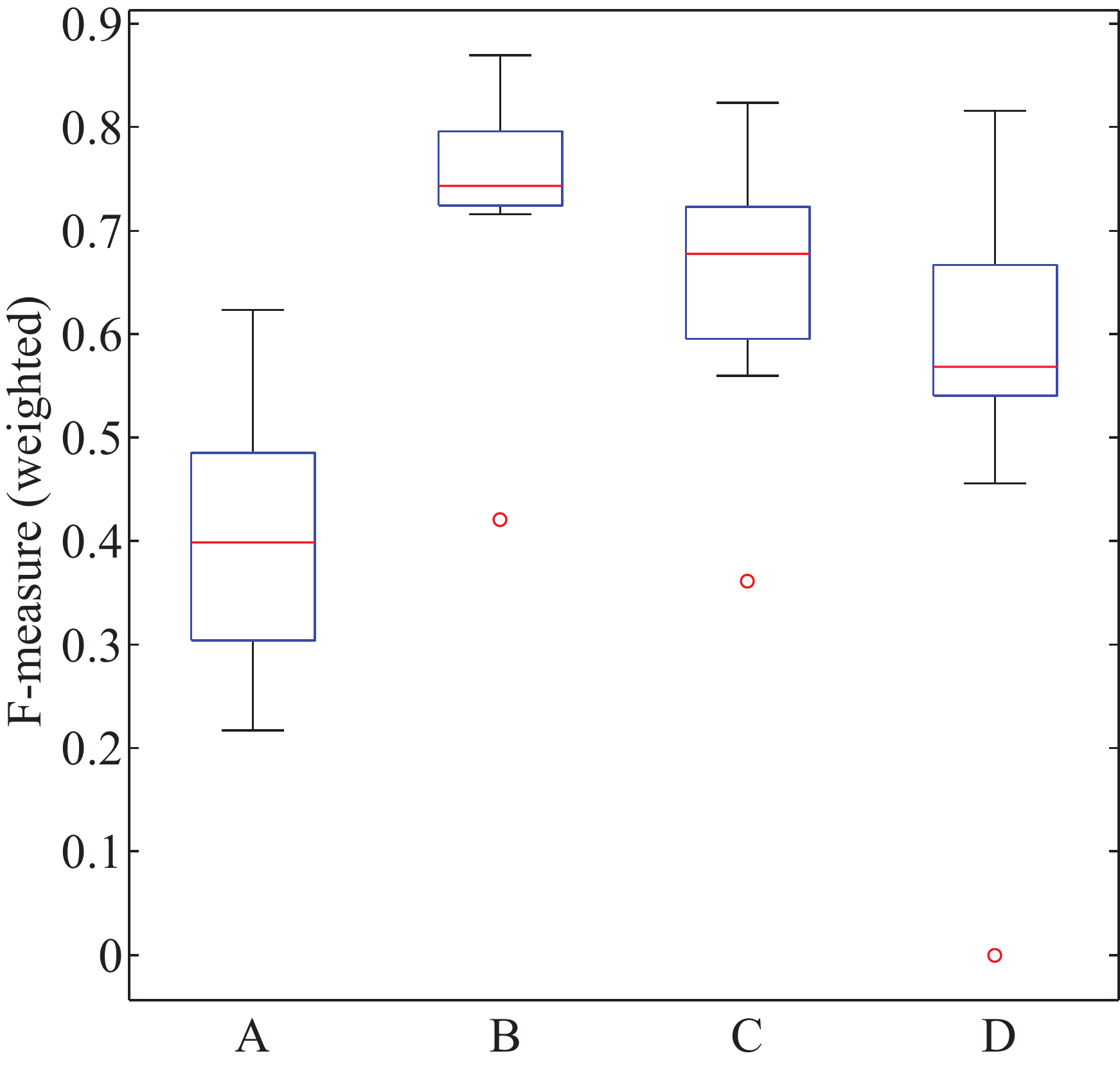}
	\qquad\qquad
	\includegraphics[height=.35\textwidth]{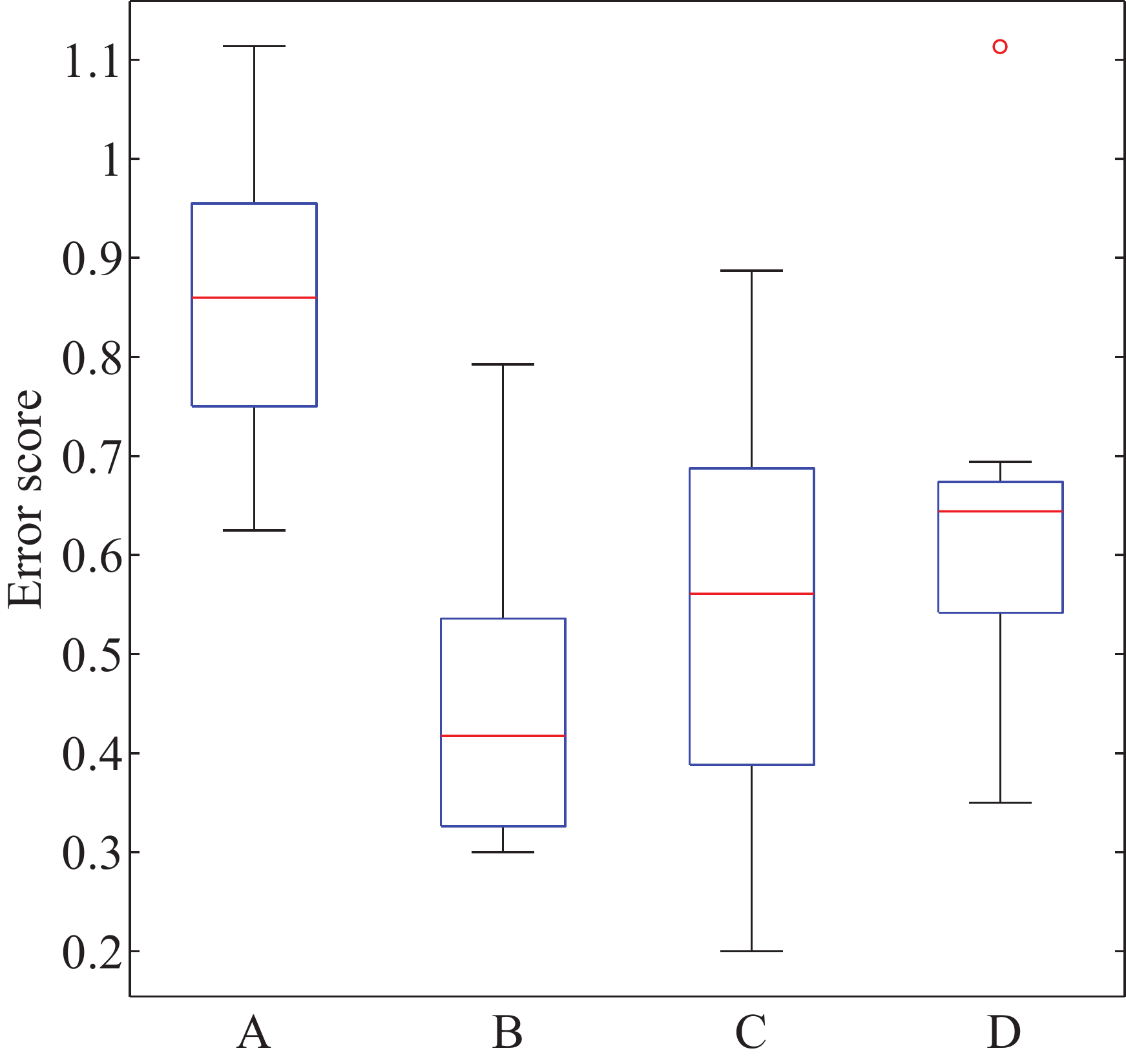}
	\label{fig:decay}}
\caption{Comparison of the systems under test w.r.t.\ the F-measure and the error score.}
\label{fig:results_1}
\end{figure*}

The results of our experiment are summarized in Fig.~\ref{fig:results_1}. To assess the algorithms' accuracy on how well they manage to detect the correct pitch value as a function of time, we use for the time grid the duration of a thirty-second note as the step size. This applies to Fig.~\ref{fig:pitch}. For the onset detection, a grid resolution equal to a sixteenth note is chosen. This is because we can be more tolerant to onset deviations in view of the fact that the onsets in the reference transcription may not always be exact. With that said, we can afford to be even more permissive with the note durations, since the decay of a note is not always easily defined. Therefore, we choose a time grid with a resolution of an eighth note in Fig.~\ref{fig:decay}. On each box, the central mark is the median and the edges of the box are the 25th and 75th percentiles. The whiskers extend to the most extreme data points. The outliers are plotted individually as circles. 

From Fig.~\ref{fig:pitch} it can be seen that the proposed transcriber \textcolor{blue}{outperforms} the \textcolor{blue}{reference transcriber}\textcolor{red}{, which performs worse\footnote{\color{red} This is partly due to the more sophisticated pitch tracking in PYIN.} than PYIN, and also PYIN} in terms \textcolor{blue}{of pitch} detection ability. \textcolor{red}{The pitch tracker stabilizes and further improves the accuracy of the pitchgram by some few percent: compared to Fig.~\ref{fig:cmp_nmf}, the box of \emph{C} is narrower and the median moved up the scale (from 0.78 to 0.80).} This is the case for both variants of the pitchgram, although the frequency-domain pitchgram yields somewhat better results. This is explained by the following. 
\begin{itemize} 
\item The time-pitch distribution is weighted \emph{prior to} bident filtering in the case of \emph{B}\textcolor{blue}{.} 
As a consequence, we observe less octave ambiguities. 
\item The \emph{implicit} detection of transients \textcolor{blue}{works way better for the frequency-domain pitchgram.} 
Consequently, tremolos may not be detected equally well. 
\item The transcriber's empirical thresholds were tuned to the \textcolor{blue}{frequency}-domain pitchgram in the first place. The two representations are equivalent but not identical. 
\end{itemize} 
The improvement from \emph{A} to \emph{B} is around 34 \% in terms of the F-measure and it is \textcolor{blue}{52 \%} w.r.t.\ the error score. \textcolor{red}{Respectively, the improvements from \emph{D} to \emph{A} are 19 \% and 45 \%.} Figs.~\ref{fig:onset} and \ref{fig:decay} also show an improvement. This means that the note onset and the duration can be detected more reliably using the pitchgram. The exact performance figures for each algorithm show that the outliers stem from the third excerpt that contains a progression of chords. \textcolor{blue}{We would like to highlight} that \textcolor{red}{both} \\the reference transcriber that uses YIN \textcolor{red}{and PYIN} \textcolor{blue}{fail} almost completely on that test item, \textcolor{blue}{while} our pitchgram transcriber achieves an F-score of 0.45. Two-thirds of the notes were left undetected, however. 

\begin{figure*}[!t]
\subfiguretopcaptrue
\centering
\subfigure[Substitutions: onset (left) and decay (right)]{
	\includegraphics[height=.35\textwidth]{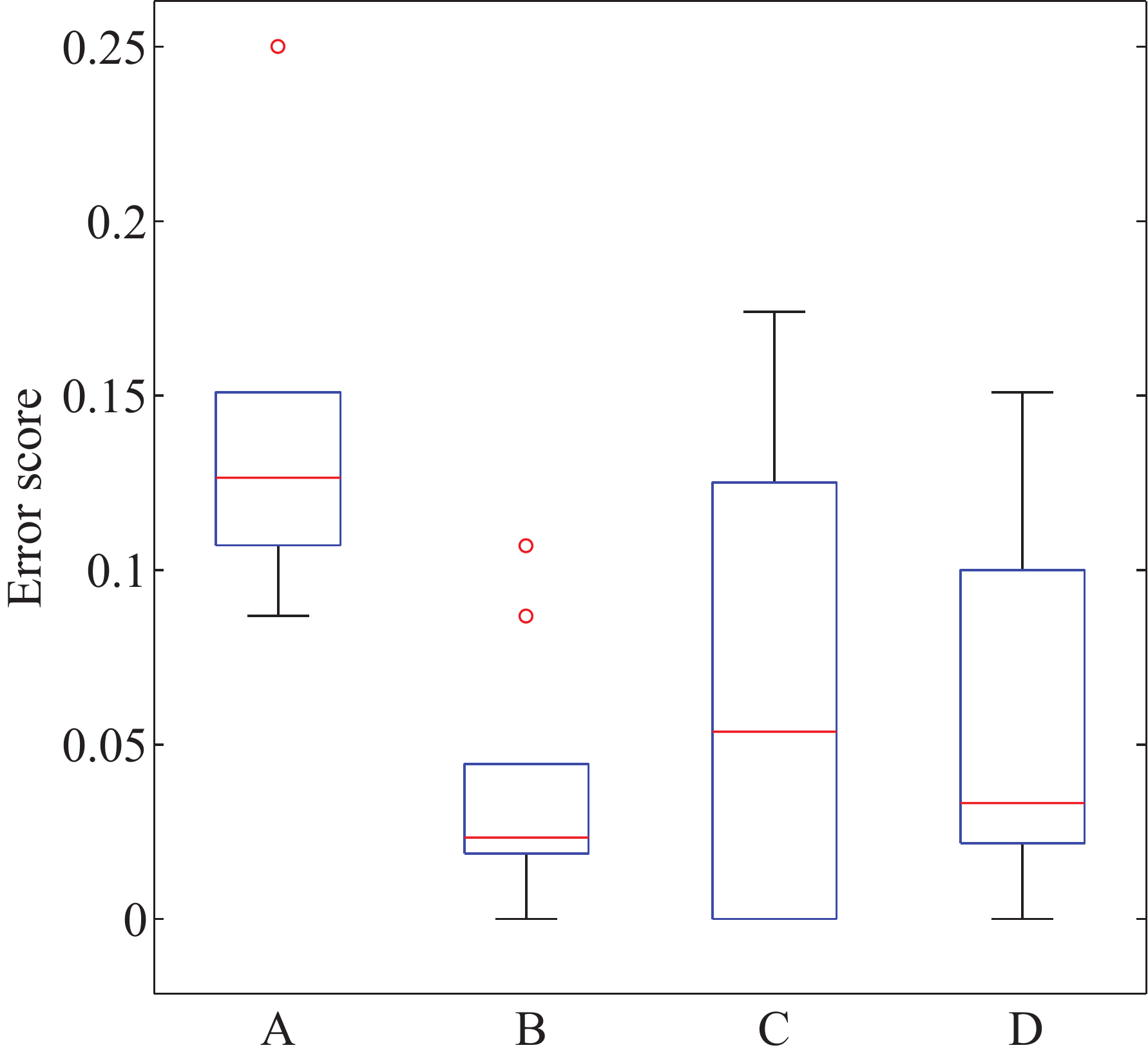}
	\qquad\qquad
	\includegraphics[height=.35\textwidth]{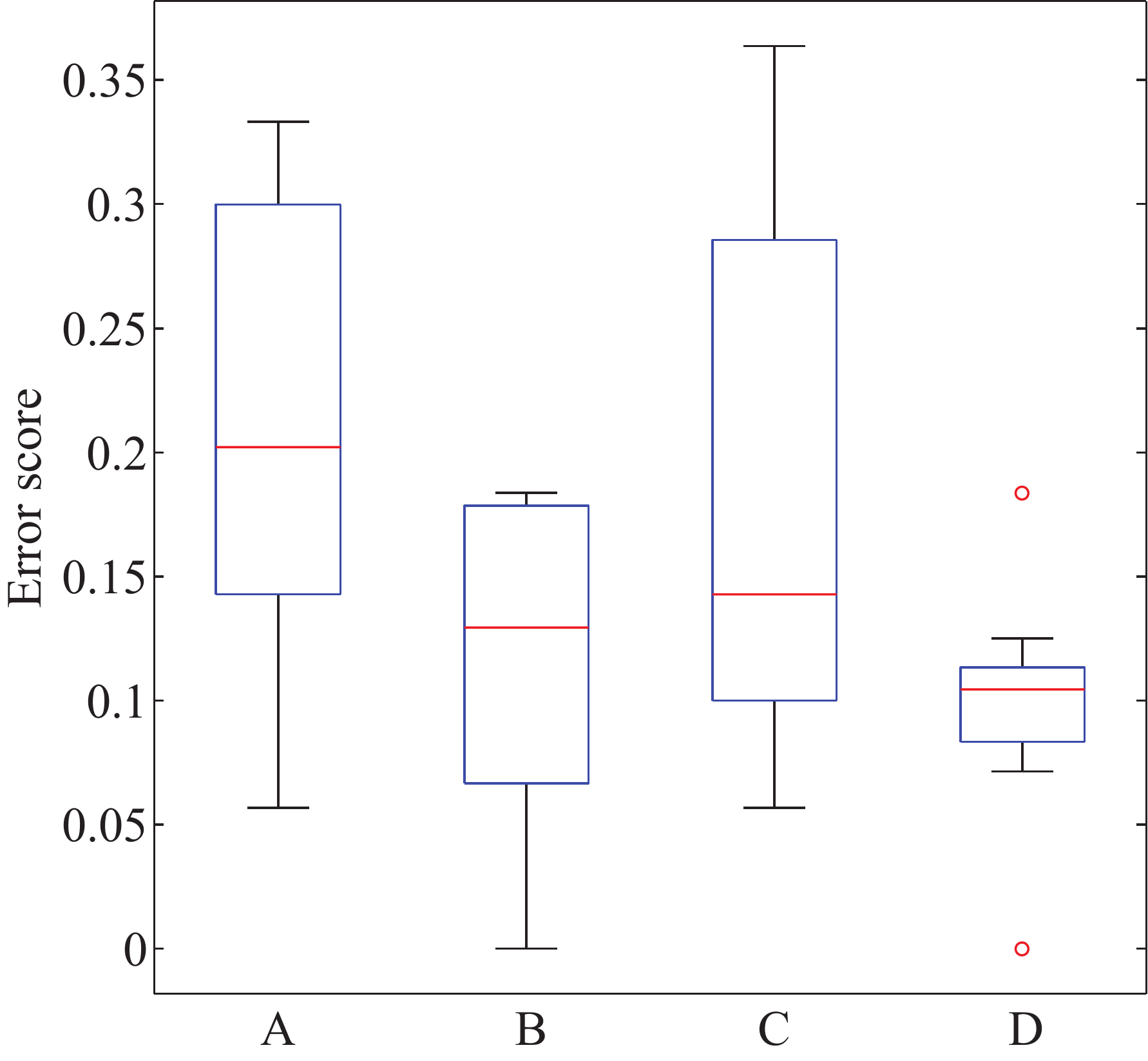}
	\label{fig:sub}}
\\
\subfigure[Deletions: onset (left) and decay (right)]{
	\includegraphics[height=.35\textwidth]{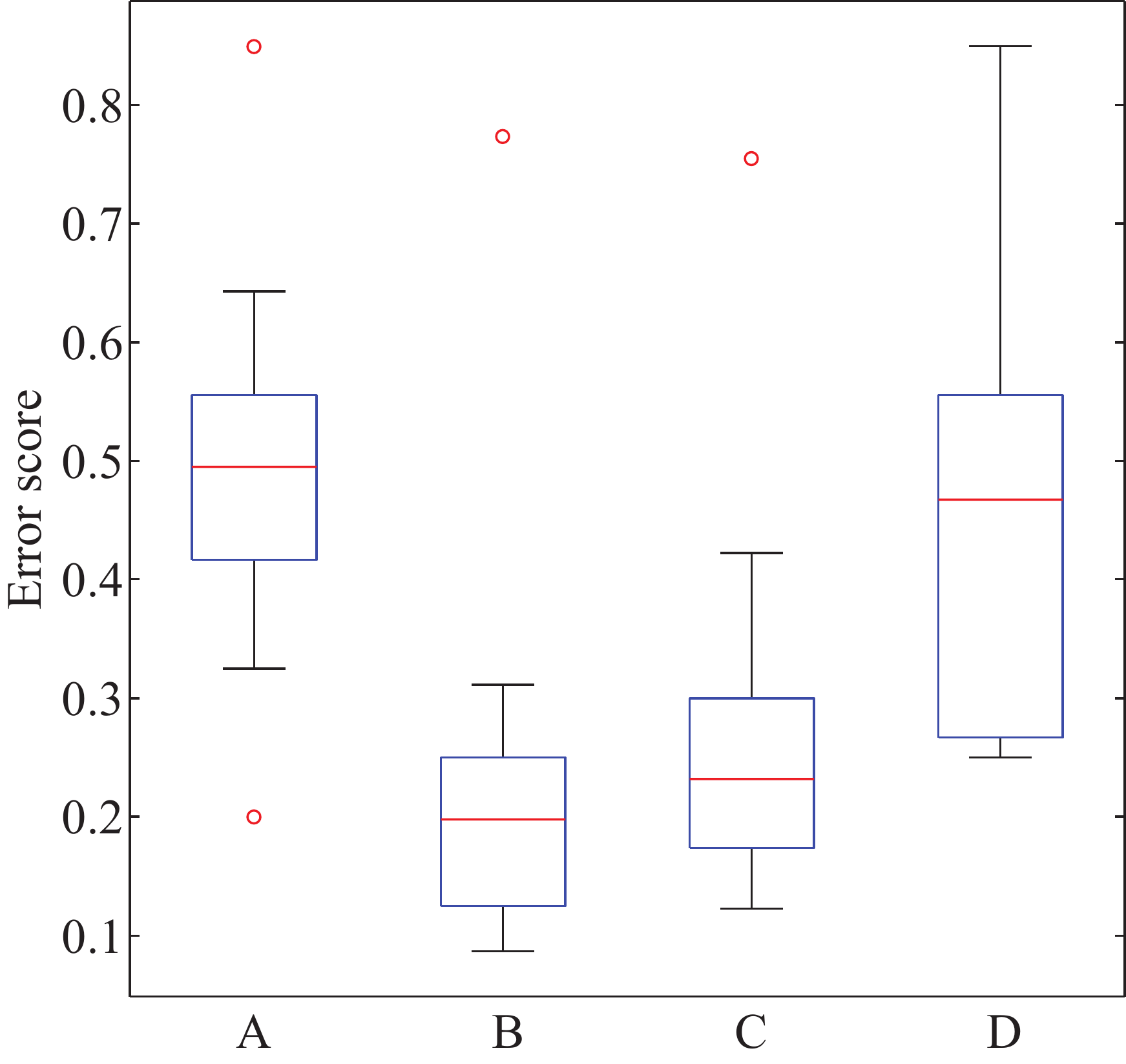}
	\qquad\qquad
	\includegraphics[height=.35\textwidth]{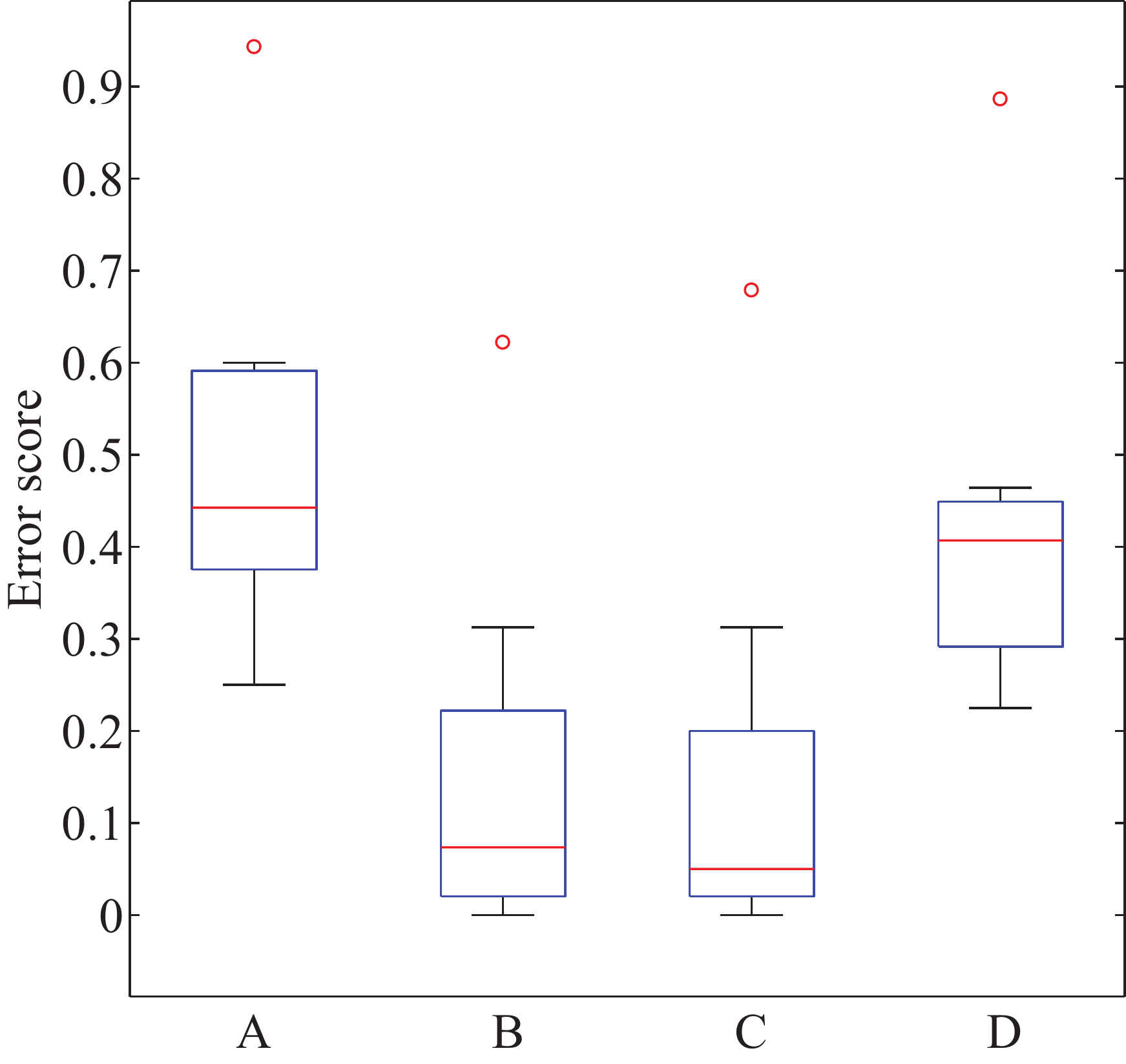}
	\label{fig:del}}
\\
\subfigure[Insertions: onset (left) and decay (right)]{
	\includegraphics[height=.35\textwidth]{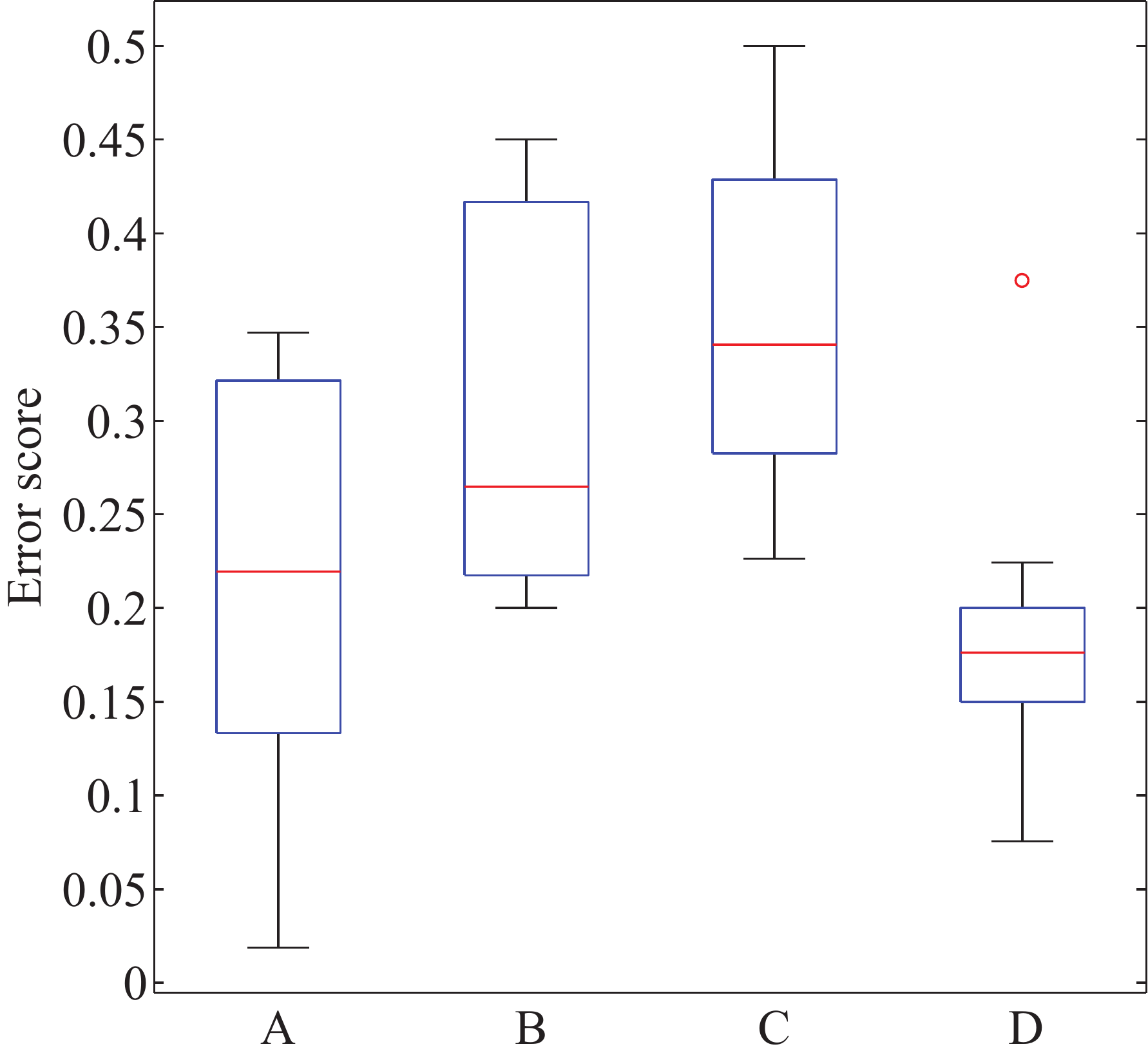}
	\qquad\qquad
	\includegraphics[height=.35\textwidth]{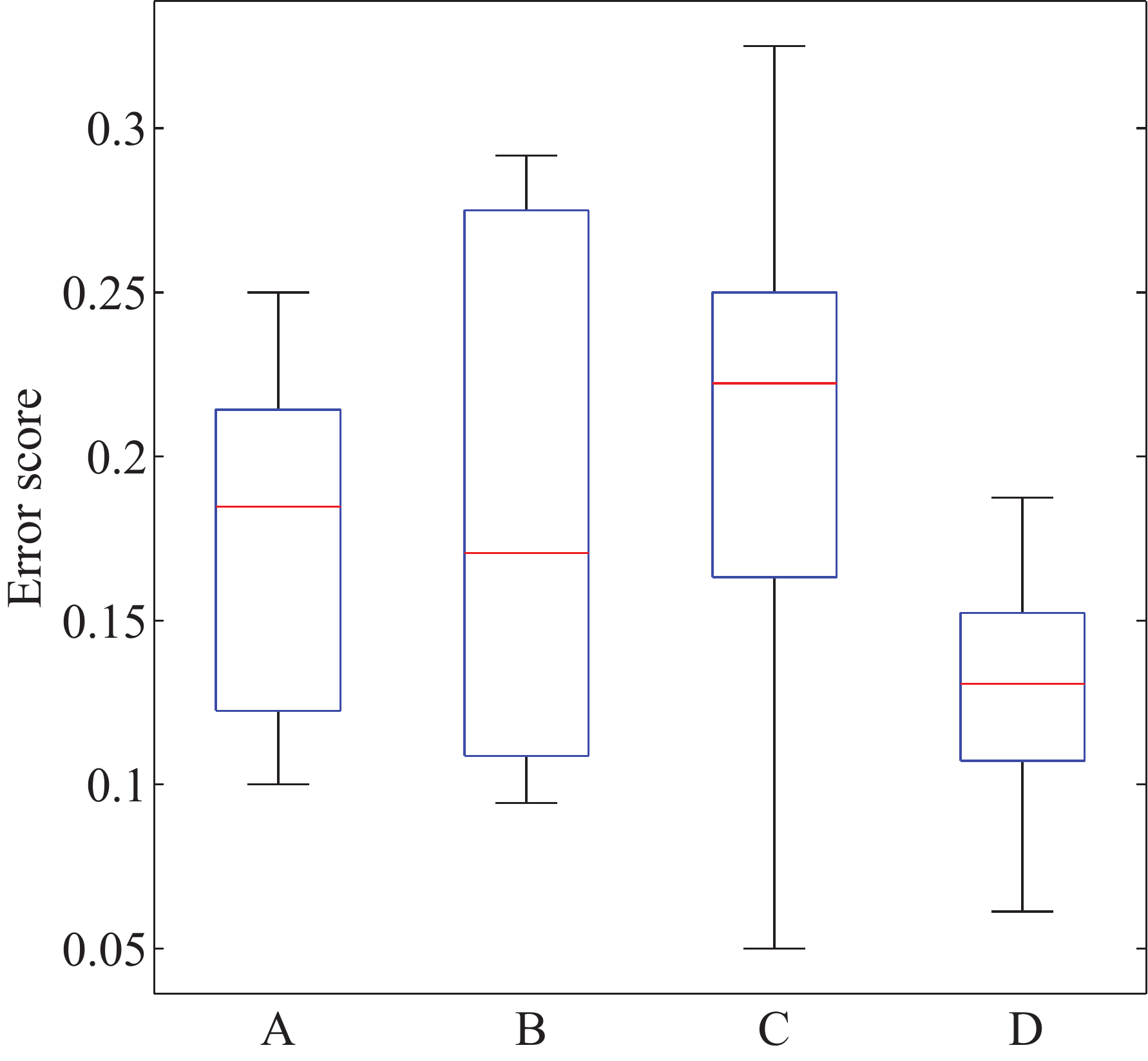}
	\label{fig:ins}}
\caption{Composition of the error score at transition points between notes.}
\label{fig:results_2}
\end{figure*}

Finally, to shed some light on the behavior of each tested system at the transition points between consecutive notes and to better understand their weak spots, we can analyze what the error score is actually composed of. This is depicted in Fig.~\ref{fig:results_2}. It can be seen that the main source of error for the reference transcriber are deletions. In part, this is an indicator of an imprecise onset detector. Also, the number of substitutions (pitch errors) near note onsets is significantly higher. \textcolor{red}{PYIN, or rather its pitch tracker, tries to detect the dominant pitch, and hence it suffers mainly from deletions.} Our system, then again, shows more insertions. I.e.\ it has detected more notes than there are according to the ground truth. These extra notes reflect string vibrations triggered by finger movements or other mechanical phenomena that make up the nuances of a live performance. For this, one could also argue that our system is better suited for live performance analysis. 

\textcolor{red}{The runtime\footnote{\color{red}Measured on an Intel Core i7-4510U CPU operating at 2.00 GHz.} of our two systems \emph{B} and \emph{C} relative to the real time, i.e.\ the total signal duration, is given in Table~\ref{tab:runtime}. It can be seen that in \emph{B} the pitchgram computation requires less than one tenth of the time to compute the regular pitchgram. The complexity of the transcriber is negligible. Moreover, \emph{B} runs four times faster than real time.} 

\begin{table}[!t]
\color{red}
\renewcommand{\arraystretch}{1.1}
\caption{Real-Time Factor of the Proposed Transcription System}
\label{tab:runtime}
\centering
\begin{tabular}{lcc}
\toprule
& B & C \\
\midrule
Frequency-domain pitchgram & 0.22 & --- \\
Time-domain pitchgram & --- & 3.02 \\
\midrule
Incl.\ transcriber & 0.25 & 3.05 \\
\bottomrule
\end{tabular}
\end{table}

\section{Conclusion}
\label{sec:conclusion}

The filter bank developed in this \textcolor{blue}{manuscript} enables a pitch-synchronous analysis of music signals. The output of the filter bank, the pitchgram, is equivalent to the real Fourier transform of a pre- and post-filtered ACF of a (\textcolor{blue}{wide-sense}) periodic signal. Therefore, the behavior of the filter bank can be analyzed based on \textcolor{blue}{the classical} theory on linear time-invariant systems. What is more, it renders possible the design and analysis of new and eventually more performant prototype filters. The bident is an example of a filter that can help alleviate the pitch ambiguity between a harmonic and its overtones. The \textcolor{red}{proposed} pitchgram may not only be useful for transcription purposes\textcolor{red}{. Eventually,} it may \textcolor{blue}{also} provide the basis for statements on the expressiveness and richness of a performance, and its technical accuracy alike\textcolor{red}{, see \cite{Jure2012}}.

An automatic, decision-based transcription \textcolor{blue}{system} for the pitchgram was also developed. It was validated that the \textcolor{blue}{system} is capable of converting a jazz guitar \textcolor{blue}{solo} into a piano roll very accurately, while taking \textcolor{blue}{certain} subtleties \textcolor{red}{(such as a high note rate, fretting noise, and pitch modulations)} of a (live) performance into account. Therefore, apart from the visual examination of the pitchgram, the obtained transcription \textcolor{blue}{could} \textcolor{red}{already} help carry out a basic yet robust performance analysis by means of a learning agent. Moreover, it should also be technically feasible to detect various playing techniques\textcolor{red}{, which is yet to be shown}. This is an outlook on \textcolor{blue}{future} work. 

The transcriber operates on an ad-hoc basis, i.e.\ without previous training. If combined with frequency-domain pitchgram, the transcriber has the potential to be run on-line and in real time on a low-power DSP. The faster pitchgram variant does not require an additional transient detector to identify tremolo or legato. In this all-in-one solution, everything is derived from the pitchgram. On the other hand, the time-domain pitchgram can provide cleaner tonality estimates at low frequencies in particular, with less frequency spread and blur. This is achieved with a bigger computational effort and with additional memory consumption. 

\textcolor{red}{As a last point, we would like to underline that the major gain of our system stems from the signal representation and not so much from the pitch tracker. As the evaluation shows, with a robust signal representation and a rather simple pitch tracker, we were able to achieve a better result than the YIN-\\based systems, despite a transient detector or a hidden Markov model with Viterbi decoding in the post-processing stage.}

\section*{Acknowledgment}

The jazz guitar licks were performed by Fran\c{c}ois Pachet. The authors cordially thank Josep~L.~Arcos from the ``Institut d'Investigaci\'o en Intel\textperiodcentered lig\`encia Artificial'' (IIIA) %
in Bellaterra %
for the provided support. %
They also thank all the reviewers of the manuscript. 

\ifCLASSOPTIONcaptionsoff
  \newpage
\fi



\bibliographystyle{IEEEtran}
\bibliography{references}

%








\end{document}